\documentclass[preprint,12pt,number]{elsarticle}
\usepackage{amssymb}
\usepackage{amsmath}
\usepackage{graphicx}
\usepackage{amsmath}
\usepackage{bm}
\usepackage[version=4]{mhchem}
\usepackage{siunitx}
\usepackage{longtable,tabularx}

\usepackage{subcaption}
\usepackage{multirow}
\usepackage{float}
\floatstyle{plain}
 
\usepackage{tikz, pgfplots}
\usetikzlibrary{angles, quotes}
\usetikzlibrary{arrows.meta, positioning, fit, calc, chains, shapes}
\usepackage{relsize}
\usepackage{xcolor}
\usepackage{colortbl}
\usepackage{array}
\usepackage{csvsimple}
\usepackage{diagbox}
\usepackage{booktabs}
\usepackage{makecell}
\usepackage{titlesec}

\DeclareMathOperator*{\argmin}{argmin}

\definecolor{label_color}{RGB}{244 , 244, 244}

\definecolor{pi_mile}{RGB}{214, 219, 212}
\definecolor{diploma}{RGB}{249, 246, 229}
\definecolor{bold_blue}{RGB}{58, 93, 174}
\definecolor{olympic_teal}{RGB}{0, 140, 149}
\definecolor{electric_blue}{RGB}{100 204 201}
\definecolor{canopy_lime}{RGB}{164, 210, 51}
\definecolor{new_horizon}{RGB}{224, 79, 57}

\titleformat{\subsection}[block]
{\normalfont\bfseries}  % This makes subsection titles bold
{\thesubsection}{1em}{}{}

\journal{Journal of Computational Physics}

\begin{document}

\begin{frontmatter}



\title{Physics-Informed Field Inversion for Sparse Data Assimilation} %% Article title


\author{Levent Ugur and Beckett Y. Zhou} %% Author name

\affiliation{organization={Georgia Institute of Technology},%Department and Organization
            city={Atlanta},
            postcode={30332-0150}, 
            state={GA},
            country={USA}}

%% Abstract
\begin{abstract}
Data-driven methods keep increasing their popularity in engineering applications, given the developments in data analysis techniques. Some of these approaches, such as Field Inversion Machine Learning (FIML), suggest correcting low-fidelity models by leveraging available observations of the problem. However, the solely data-driven field inversion stage of the method generally requires dense observations that limit the usage of sparse data. In this study, we propose a physical loss term addition to the field inversion stage of the FIML technique similar to the physics-informed machine learning applications. This addition embeds the complex physics of the problem into the low-fidelity model, which allows for obtaining dense gradient information for every correction parameter and acts as an adaptive regularization term improving inversion accuracy. The proposed Physics-Informed Field Inversion approach is tested using three different examples and highlights that incorporating physical loss can enhance the reconstruction performance for limited data cases, such as sparse, truncated, and noisy observations. Additionally, this modification enables us to obtain accurate posterior correction parameter distribution with limited realizations, making it data-efficient. The increase in the computational cost caused by the physical loss calculation is at an acceptable level given the relaxed grid and numerical scheme requirements.

\end{abstract}

\begin{graphicalabstract}
% \begin{figure}[htbp]
% \begin{center}
\resizebox{\textwidth}{!}{
    \begin{tikzpicture}[align=center,
    general_style/.style={rectangle, draw=black!100, thick, minimum width=80, minimum height=45, font=\sffamily},
    arrow/.style={->, ultra thick, black, >=Latex , font=\sffamily}]

    % \definecolor{bold_blue}{RGB}{58, 93, 174}
    % \definecolor{olympic_teal}{RGB}{0, 140, 149}
    % \definecolor{electric_blue}{RGB}{100 204 201}
    % \definecolor{canopy_lime}{RGB}{164, 210, 51}
    % \definecolor{new_horizon}{RGB}{224, 79, 57}

    \node[general_style, fill=bold_blue!50] (optimizer) {Optimizer \\ ($\argmin_{\beta}J$)};

    \node[general_style, fill=olympic_teal!50] (lofi) [right of=optimizer, xshift=4.0cm] {Low-Fidelity \\ Model ($\mathcal{H}_\beta$)};

    \node[general_style, fill=electric_blue!50] (pred) [right of=lofi, xshift=4.0cm] {Predictions \\ ($\mathcal{H}_\beta(\mathbf{x})$)};

    \node[general_style, fill=new_horizon!50] (data_loss) [above of=pred, yshift=1.5cm] {Data Loss \\ ($J_{data}$)};

    \node[general_style, fill=canopy_lime!50] (hifi_data) [right of=data_loss, xshift=3.0cm] {Observations \\ ($\mathcal{M}(\mathbf{x})$)};

    \node[general_style, fill=new_horizon!50, dashed] (phys_loss) [below of=pred, yshift=-1.5cm] {Physical Loss \\ ($J_{phys}$)};

    \draw[arrow]	(optimizer) -- (lofi) node[midway, above] {\textit{corrects}};
    \draw[arrow]	(lofi) -- (pred) node[midway, above] {\textit{predicts}};
    \draw[arrow]	(pred) -- (data_loss);
    \draw[arrow, dashed]	(pred) -- (phys_loss);
    \draw[arrow]	(hifi_data) -- (data_loss);
    \draw[arrow]	(data_loss) -| (optimizer) node[pos=0.36, above] {\large{Data-Driven Inversion}};
    \draw[arrow, dashed]	(phys_loss) -| (optimizer) node[pos=0.332, below] {\large{Physics-Informed Inversion}};
    
    \end{tikzpicture}
    }
% \end{center}
% \caption{Proposed Physics Informed Field Inversion Framework. The upper loop illustrates the conventional data-driven field inversion process, while the bottom loop with dashed lines represents the addition of the proposed physical loss.} \label{wf:framework}
% \end{figure}
\end{graphicalabstract}

\begin{highlights}
\item A physical loss term is introduced into the field inversion stage of data assimilation to obtain appropriate correction parameters for low-fidelity models.
\item The proposed loss term enables recovering the properties across the solution space for various challenging inversion scenarios with sparse observations by incorporating both available data and the physical background of the problem.
\item The physical loss addition provides more accurate posterior correction parameter statistics for single realization cases compared with solely data-driven inversion, making the process more data-efficient.
\end{highlights}

%% Keywords

\end{frontmatter}


% \title{Physics Informed Field Inversion for Data Assimilation}

% \author{Levent Ugur\footnote{Ph.D. Candidate, Department of Aerospace Engineering, University of Bristol, levent.ugur@bristol.ac.uk}, Filipi T. Kunz\footnote{Ph.D. Candidate, Department of Aerospace Engineering, University of Bristol, filipi.teixeirakunz@bristol.ac.uk}, and Beckett Y. Zhou\footnote{Lecturer in Aeroacoustics, Department of Aerospace Engineering, University of Bristol, beckett.zhou@bristol.ac.uk}}
% \affil{University of Bristol, Bristol, BS8 1TH, United Kingdom}

% \begin{document}

% \maketitle

% \begin{abstract}
%     Abstract
% \end{abstract}

\section{Introduction}
\label{sec:introduction}

% One of the approaches of data assimilation in CFD focuses on correcting the tuning parameters of turbulence closure models \cite{} as global constants \cite{}. However, this approach has a low degree of correction freedom and generally the improvement is whether limited or case-specific \cite{}. To increase the control over the low-fidelity model prediction, more advanced data assimilation techniques are employed. 

% Need of data assimiliaton
Resource limitations have been one of the biggest challenges in engineering design studies for decades. Despite developments in hardware technologies making the high-fidelity computational tools more accessible, the demands of the industry are increasing even more rapidly. This means that the resources available per design stage remain highly restricted, even though the total number of resources is increasing. Therefore, it is always desirable to achieve results with high accuracy by using tools with lower costs. Consequently, it is necessary to develop innovative approaches to utilize available resources efficiently, as stated in the Vision 2030 Study by NASA \cite{Slotnick_2014}.

One of the most valuable resources in the big data age is the knowledge obtained from similar engineering applications. Developments in computational simulations and advanced measurement techniques make it possible to collect a large volume of data in engineering applications \cite{Brunton_2021}. One of the best ways to transfer existing knowledge to new computational simulations is using data assimilation techniques \cite{Lewis_2006}, which allows us to predict the results in the quality of high-quality observations with a lower cost for similar engineering problems, which is highly desirable for the frameworks where similar problems are repeatedly analyzed such as design optimization processes.

% Data assimilation
There are two main types of data assimilation techniques. The first type is sequential techniques, such as the Kalman Filter \cite{Kalman_1960} and Ensemble Kalman Filter \cite{Lakshmivarahan_2009}, focusing on correcting the state vectors. The second type is variational techniques \cite{LeDimet_1986, Rabier_2003}. These techniques focus on correcting the model parameters or introducing forcing terms in the equations, either spatially or temporally, over a given window interval or for averaged values \cite{Hayase_2015}. The aim is to minimize the loss between the predictions obtained from low-fidelity models and the high-fidelity observations to optimize the correction parameters of the low-fidelity model. There are some studies comparing the performance of sequential and variational techniques. Studies by Mons et al. \cite{Mons_2016} and Alvarado-Montero et al. \cite{Alvarado-Montero_2022} showed that variational techniques outperform sequential ones in various cases. Despite the accuracy advantage of variational techniques, their main disadvantage is their cost, as they are gradient-based iterative methods \cite{Hayase_2015}. However, recent developments in computational techniques for calculating derivatives, such as automatic differentiation \cite{auto_diff}, enable fast, accurate, and mesh-free computation of adjoint sensitivities for large-scale vectors. These developments encourage researchers to employ variational data assimilation in large-scale engineering applications as well.

% Generalizability of the DA and FIML 
A question that arises with the application of data assimilation is its generalizability and the applicability of the knowledge obtained from one assimilation case to other cases. Some studies in the literature \cite{He_2019, Ozawa_2024} use data assimilation as the first step in advanced turbulence investigation frameworks, such as resolvent mode analysis \cite{McKeon_2010}. This approach employs data assimilation to recover an accurate flow field with higher resolution, which can then be used in resolvent analysis to better understand the underlying physics of the turbulent flow case under investigation. Another development to improve the generalizability of data assimilation is the Field Inversion Machine Learning (FIML) framework \cite{Parish_2016}. This approach combines variational data assimilation with machine learning, allowing the information obtained during the assimilation stage to be applied to other similar cases through the use of trained machine learning models. Although the FIML technique was primarily developed and is mostly applied to turbulent closure models such as $k$-$\epsilon$ \cite{Sanhueza_2023}, $k$-$\omega$ \cite{Ho_2021, Bidar_2022}, and Spalart-Allmaras \cite{Suarez_2021, Ferrero_2020, Duraisamy_2015}, its applications are not limited to these cases. Recent studies \cite{Ugur_2024, Kunz_2024} have used this technique to enhance the predictive capability of turbulence generation models for aerodynamic noise analyses. Ugur et al. \cite{Ugur_2024} demonstrated that by using machine learning models trained with various jet flow cases, the prediction performance of a low-fidelity noise generation model can be improved even for cases that were not seen during training. The applicability of FIML to unseen cases highlights the generalizability of the framework and demonstrates the efficient use of information obtained through data assimilation.

% Limitation of conventional FI
While FIML offers a powerful data assimilation methodology with a high degree of correction freedom, it still has some limitations. Firstly, the conventional approach is based on the Bayesian inversion which utilizes the statistical correlations as prior knowledge. However, this knowledge is solely data-driven and does not account for any physical considerations that align with the nature of the problem, although it is known that adding physical restrictions to limit the design space can positively impact the flow inversion processes \cite{He_2020}. Furthermore, effective Bayesian inversion requires the computation of a full covariance matrix calculation. Although this step is not particularly challenging, it necessitates a relatively dense dataset. While this can be achievable for time-averaged simulations, it may not be possible to utilize sparse experimental data as the high-fidelity data source. Even for computational observations, if the investigation of an unsteady physical phenomenon is desired, such as turbulent flow field modeling, it is not feasible to record all unsteady data generated during the analysis. The most common approach is recording the time series data at specific probe locations in the solution domain, akin to experimental setups.

% ONERA team story
Some previous studies on sparse data assimilation, particularly focused on fluid mechanics, aimed to improve low-fidelity model results with high spatial resolution by using coarse or truncated high-fidelity measurements. Some of them utilize sequential techniques such as state observer assimilation \cite{Marquet_2023}, or nudging \cite{Ishikawa_1996, Zauner_2022}. However, the majority of these studies focus on variational data assimilation methods.

Foures et al. \cite{Foures_2014} applied a forcing term approach to correct the Navier-Stokes equations for a low Reynolds number flow over a cylinder. They compared their adjoint-based variational approach using full observations for optimal assimilation, regularly reduced sparse observations to assess interpolation, and truncated observation spaces to assess extrapolation capabilities. The reconstruction results generally show good agreement with high-fidelity simulation results.

Symon et al. \cite{Symon_2017} employed a similar approach to that of \cite{Foures_2014}, making minor modifications to apply their method to a real-life engineering scenario. They studied the flow over an idealized airfoil using measurements obtained from PIV data, which has a coarser resolution compared to numerical results. When experimenting with different sparsity levels of measurements, they observed a decrease in accuracy and convergence issues as the measurements became coarser. This creates an uncertainty for the necessary measurement resolution, as it can limit the generalizability of data assimilation driven by experimental data.

Franceschini \cite{Franceschini_2020} expanded on previous studies by considering a backward-facing step case at higher Reynolds numbers and using Spalart-Allmaras turbulence modeling. In this study, station-based observations were used, akin to pressure measurement experiments, rather than regularly distributed sparse measurements. Their reconstruction results with sparse data exhibited non-smooth, unphysical behavior in the reconstructed field, similar to the findings in Symon et al. \cite{Symon_2017}. To address this issue, they modified their objective function by adding a forcing term gradient. They found that this addition resulted in a smoother reconstruction, acting as a form of physical-like regularization. The experience obtained from this study highlights the importance of incorporating physical effects alongside data-driven assimilation approaches. This insight guided Patel et al. \cite{Patel_2023} to employ Physics-Informed Neural Networks (PINNs) \cite{Raissi_2019} to enhance the reconstruction results achieved by Franceschini \cite{Franceschini_2020}. They demonstrated that integrating physical constraints into the inversion process improves the accuracy of the reconstruction.

% PIML
Incorporating physical loss during an optimization process in addition to the data loss is a well-established approach in Physics-Informed Machine Learning (PIML) \cite{Raissi_2019, Karniadakis_2021} applications. This strategy significantly enhances the interpolation and extrapolation capabilities of machine learning models, which are known as major limitations of conventional, solely data-driven machine learning models in engineering applications \cite{Li_2022}. PIML methods replace these black-box machine learning models with their interpretable physics-aware counterparts. Moreover, employing the machine learning frameworks to train universal approximators \cite{hornik1989multilayer}, such as neural networks, allows computation of derivatives via backpropagation enabling straightforward and mesh-free derivatives.

For these reasons, PIML is a popular approach for property reconstruction across various engineering disciplines \cite{Patel_2023, He_2020, Bai_2020}. However, these frameworks utilize machine learning models for the forward modeling, which still often lack interpretability compared to well-developed low-fidelity models based on physical relationships. One of the primary advantages of low-fidelity models is their reliability across unseen cases by offering a strong baseline, including those substantially different from the cases used in the training dataset. Additionally, Pestourie et al. \cite{Pestourie_2023} demonstrated that low-fidelity models, when corrected at the input level using machine learning, can perform better than feedforward neural networks as surrogates for limited data cases. This highlights that employing low-fidelity models not only increases the reliability of the solution but also reduces the amount of data needed to achieve accurate predictions, compared with the ones using neural networks as the universal approximators.

% Proposed approach
This paper proposes incorporating an additional physical loss in the field inversion stage of the FIML method, inspired by PIML applications. Although FIML is already classified as a hybrid physics-ML modeling approach \cite{Willard_2020}, the physics considered in the process is obtained by the low-fidelity model which may not fully capture the complete physics of the problem. Incorporating a more advanced physical loss aims to achieve two significant improvements during the inversion process. Firstly, this loss term allows us to gather the adjoint information for every parameter, even in the absence of observations at many locations. The significance of dense adjoints for improving reconstruction performance is discussed in \cite{Symon_2017}. Secondly, incorporating a physical loss is expected to enhance the performance of inverse optimization by acting as an adaptive regularization term. This term will avoid unphysical results, which are typically missing in solely data-driven approaches. The proposed inversion approach has the potential to enhance the accuracy of interpretable low-fidelity models by leveraging both available data and the physical background of the problem.

% Organization
The rest of the paper is organized as follows: the methodology of the Physics-Informed Field Inversion (PIFI) approach is introduced in Section \ref{sec:methodology}. To showcase how beneficial it can be in different field inversion scenarios, PIFI is tested on three different sample cases ordered as exponential growth, 1D heat conduction, and turbulent velocity field modeling. The results of these test cases are presented in Section \ref{sec:application}. Finally, the key outcomes of the study are summarized in Section \ref{sec:conclusion}, along with a discussion of the proposed approach, including its achievements and limitations.

\section{Methodology}
\label{sec:methodology}

Field Inversion Machine Learning (FIML) is an advanced data assimilation technique that combines a data-driven field inversion process with machine learning strategies to enable enhanced low-fidelity model predictions across a variety of similar cases.

The methodology of FIML is introduced in the study by  Parish and Duraisamy \cite{Parish_2016}. The approach uses a two-step process to train machine learning models that can later be used in predictive mode. These models are utilized to create a mapping between the low-fidelity model inputs and the model parameters such that, for every point in the solution domain, we can predict the most appropriate correction coefficients to be applied to the model parameters. This approach allows us to employ input-dependent and spatially varying model parameters, which is highly desirable for most low-fidelity models, as their empirical constants are generally one of the reasons for their lower accuracies.

The first step of FIML is the field inversion stage, where the goal is to obtain the optimal correction coefficients by minimizing the discrepancy between the low-fidelity model prediction and the high-fidelity data (observations, measurements). After completing this inverse optimization step for the available cases, a dataset is obtained where the features correspond to the inputs of the low-fidelity model, while the labels contain the optimal correction coefficients.

The second step involves training machine learning models using the generated dataset. This is simply a supervised regression problem, which allows for the use of well-known machine learning models. Once trained, the machine learning model can apply the insights gained from the training cases to new, unseen cases in the predictive mode. This stage of FIML enables an effective framework to generalize the knowledge gained from data assimilation.

The traditional FIML approach uses Bayesian inversion in the field inversion step. This approach enables us to utilize prior knowledge of the data, if it is available, leading to a more robust inversion. However, due to its higher computational cost compared with the ordinary least-square method, \textit{maximum a posteriori} (MAP) approximation is often used to determine the optimal distribution of correction coefficients. The data loss component of the objective function to be minimized is defined as the discrepancy between the low-fidelity model predictions and the observations, as shown in Equation \ref{eq:data_loss}.

\begin{equation}
    J_\text{data} = \frac{1}{2}  \left( \mathcal{H}_{\beta}(\mathbf{x}) - \mathcal{M}(\mathbf{x}) \right)^T \mathbf{C_m}^{-1} \left( \mathcal{H}_{\beta}(\mathbf{x}) - \mathcal{M}(\mathbf{x}) \right)
    \label{eq:data_loss}
\end{equation}

In Equation \ref{eq:data_loss}, the high-fidelity data source function is denoted as $\mathcal{M}$, which is a function of the physical input state, $\mathbf{x}$. This high-fidelity function is not necessarily a computational simulation in this context, it can also represent measurements obtained from an experiment. Another function in this equation is the low-fidelity model, $\mathcal{H}$, which depends on both $\mathbf{x}$ and $\beta$, where $\beta$ represents the correction parameters used to tune the low-fidelity model parameters. The discrepancy between these two functions is calculated using $N_m$ observation points which can be less than the total number of grid points used in the inversion stage, $N$ for the cases with sparse observations.

The matrix $\mathbf{C_m}$ in Equation \ref{eq:data_loss} represents the observational covariance matrix for data loss. Its importance is explained in \cite{Parish_2016} as a full covariance matrix calculation significantly improves the accuracy of the posterior statistics of the correction parameter distributions. However, the study also highlights that the estimation obtained from simpler covariance matrices can still be useful for recovering the solution states. In fact, for cases where there are not enough realizations, such as the inversion of a single case, calculating the full covariance matrix may not be possible. In such cases, a simpler choice for this matrix can be using a diagonal matrix defined as $\mathbf{C_m}=N_m \mathbf{I}$ which converts the data loss function to a mean squared error (MSE) format as presented in Equation \ref{eq:data_loss_mse}.

\begin{equation}
    J_\text{data} = \frac{1}{N_m} \sum_{i=1}^{N_m} \left| \mathcal{H}_{\beta}(x^i) - \mathcal{M}(x^i) \right|^2
    \label{eq:data_loss_mse}
\end{equation}

Since this data-driven inversion is inherently an ill-conditioned optimization problem \cite{Ferrero_2020}, it is necessary to apply regularization which is calculated as the discrepancy between the optimized correction values and their prior distribution, aligns with conventional FIML formulation and is presented in Equation \ref{eq:regularization}.

\begin{equation}
    J_\text{reg} = \frac{1}{2} \left( \beta - \beta_\text{prior} \right)^T \mathbf{C_\beta}^{-1} \left( \beta - \beta_\text{prior} \right)
    \label{eq:regularization}
\end{equation}

Since there is usually no direct knowledge for the covariance correction coefficients, the prior covariance matrix $\mathbf{C_\beta}$ can be similarly replaced with a simple diagonal matrix with $\mathbf{C_\beta}=N_\beta \mathbf{I}$. In this case, the total number of correction coefficients, $N_\beta$, serves as the variance, as the regularization is applied to the full set of the correction coefficients. $N_\beta$ typically scales with the number of grid points, $N$, used in the low-fidelity model calculation and determines the degree of freedom of the inverse optimization. With the assumption of diagonal covariance, the regularization loss also simplifies to the MSE loss as outlined in Equation \ref{eq:regularization_mse}.

\begin{equation}
    J_\text{reg} = \frac{1}{N_\beta} \sum_{i=1}^{N_\beta} \left| \beta^i - \beta_\text{prior}^i \right|^2
    \label{eq:regularization_mse}
\end{equation}

As we discussed, simplifying the Bayesian inversion to ridge regression is reasonable and can even be mandatory for some cases. However, such a simplification can significantly degrade the advantage obtained by leveraging the prior physical knowledge during the inversion process. This knowledge is not only advantageous for obtaining accurate posterior statistics but is also essential for optimization with sparse observations. However, given that the models of interest are physical representations, there are alternative ways of embedding physical knowledge into optimization frameworks rather than relying solely on data-driven statistics. A well-known approach for this purpose is employing a physical loss term similar to physics-informed machine learning applications. This physical loss can be combined with the data loss during the optimization stage and minimized by the optimizer. The selection of the physical loss is typically case-specific. Nonetheless, for most engineering cases, we can consider the residual of a differential equation to serve as the physical loss.

Let us define an arbitrary differential equation as $f(\mathbf{x}, \mathbf{y})=0$, where $\mathbf{x}$ is the input state and $\mathbf{y}$ is the output state of the models. Ideally, the output generated from the low-fidelity function, $\mathcal{H}(\mathbf{x})$, should satisfy this equation. However, this is not the case for most low-fidelity results. The proposed approach utilizes the residual of this differential equation as a loss term so that the optimizer will encourage the low-fidelity model to predict an output state that minimizes the residual of the equation. Consequently, this results in outputs that are more physics-aware, rather than simply matching the provided data.

Equation \ref{eq:physical_loss} outlines the general form of the physical loss. For any input set of $\mathbf{x}$, the low-fidelity model, corrected by a set of correction parameters $\beta$, is employed to predict the output states $\mathbf{y}$. Once these predictions are made, the physical system can be employed to calculate the residual, which serves as the physical loss. It is important to note that appropriate initial and boundary conditions should also be taken into account in this loss term when necessary.

\begin{equation}
    J_\text{phys} = \frac{1}{N} \sum_{i=1}^{N} \left| f(\mathbf{x}^i, \mathcal{H}_\beta(\mathbf{x}^i)) \right|
    \label{eq:physical_loss}
\end{equation}

To create a comprehensive loss function for use in the inverse optimization stage, we combine the losses defined in Equations \ref{eq:data_loss}, \ref{eq:regularization} and \ref{eq:physical_loss}. However, this combination is not always straightforward and requires weight parameters, $\alpha$, to adjust the importance of each loss component in the total loss as shown in Equation \ref{eq:total_loss}. The selection of these weighting parameters is discussed in the following sections of the paper, but they can be considered hyperparameters that should be essentially tuned for every specific case.

\begin{equation}
    J = \alpha_\text{data} J_\text{data} + \alpha_\text{reg} J_\text{reg} + \alpha_\text{phys} J_\text{phys}
    \label{eq:total_loss}
\end{equation}

By using the combined loss function given in Equation \ref{eq:total_loss}, an optimization cycle can be performed to obtain the optimal distribution of the correction coefficients, $\beta^*$ as given in Equation \ref{eq:argmin_beta}. To solve this minimization problem, a gradient descent algorithm can be employed. However, the degree of freedom, $N_\beta$, typically scales with the number of points used in the low-fidelity model, $N$. This scaling can quickly evolve the problem into a large-scale optimization task, which requires accurate and fast gradient calculation during the inverse optimization steps. While this could be a challenging stage in the framework proposed, the developments in automatic differentiation (AD) modules enable us to obtain the adjoint gradients in an easy and scale-free way. In this study, the AD capability of JAX library \cite{jax2018github} is employed during the inverse optimization stages for that purpose.

\begin{equation}
    \beta^* = \argmin_{\beta}J
    \label{eq:argmin_beta}
\end{equation}

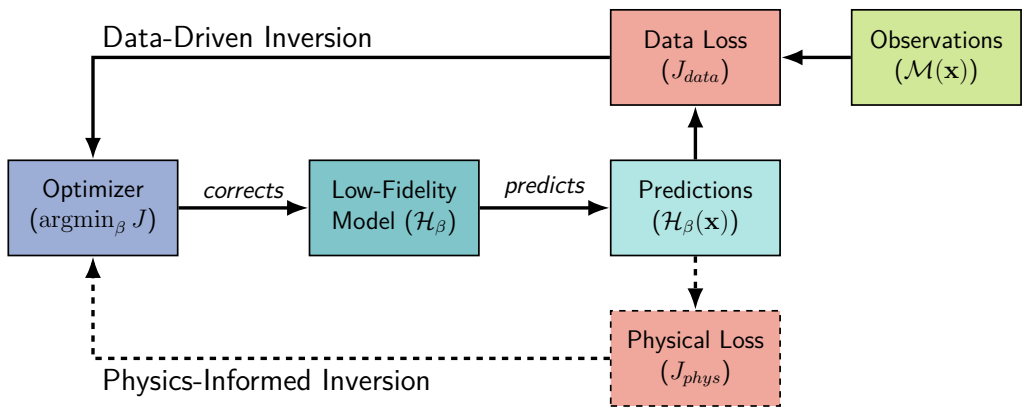
\begin{figure}[htbp]
\begin{center}
\resizebox{\textwidth}{!}{
    \begin{tikzpicture}[align=center,
    general_style/.style={rectangle, draw=black!100, thick, minimum width=80, minimum height=45, font=\sffamily},
    arrow/.style={->, ultra thick, black, >=Latex , font=\sffamily}]

    % \definecolor{bold_blue}{RGB}{58, 93, 174}
    % \definecolor{olympic_teal}{RGB}{0, 140, 149}
    % \definecolor{electric_blue}{RGB}{100 204 201}
    % \definecolor{canopy_lime}{RGB}{164, 210, 51}
    % \definecolor{new_horizon}{RGB}{224, 79, 57}

    \node[general_style, fill=bold_blue!50] (optimizer) {Optimizer \\ ($\argmin_{\beta}J$)};

    \node[general_style, fill=olympic_teal!50] (lofi) [right of=optimizer, xshift=4.0cm] {Low-Fidelity \\ Model ($\mathcal{H}_\beta$)};

    \node[general_style, fill=electric_blue!50] (pred) [right of=lofi, xshift=4.0cm] {Predictions \\ ($\mathcal{H}_\beta(\mathbf{x})$)};

    \node[general_style, fill=new_horizon!50] (data_loss) [above of=pred, yshift=1.5cm] {Data Loss \\ ($J_{data}$)};

    \node[general_style, fill=canopy_lime!50] (hifi_data) [right of=data_loss, xshift=3.0cm] {Observations \\ ($\mathcal{M}(\mathbf{x})$)};

    \node[general_style, fill=new_horizon!50, dashed] (phys_loss) [below of=pred, yshift=-1.5cm] {Physical Loss \\ ($J_{phys}$)};

    \draw[arrow]	(optimizer) -- (lofi) node[midway, above] {\textit{corrects}};
    \draw[arrow]	(lofi) -- (pred) node[midway, above] {\textit{predicts}};
    \draw[arrow]	(pred) -- (data_loss);
    \draw[arrow, dashed]	(pred) -- (phys_loss);
    \draw[arrow]	(hifi_data) -- (data_loss);
    \draw[arrow]	(data_loss) -| (optimizer) node[pos=0.36, above] {\large{Data-Driven Inversion}};
    \draw[arrow, dashed]	(phys_loss) -| (optimizer) node[pos=0.332, below] {\large{Physics-Informed Inversion}};
    
    \end{tikzpicture}
    }
\end{center}
\caption{Proposed Physics Informed Field Inversion Framework: The upper loop illustrates the conventional data-driven field inversion process, while the bottom loop with dashed lines represents the addition of the proposed physical loss.} \label{wf:framework}
\end{figure}

The proposed physics-informed field inversion framework is also summarized in Figure \ref{wf:framework}. Once this stage is completed, the supervised machine learning stage of the traditional FIML framework can be performed without any modification.

\section{Application on Sample Cases}
\label{sec:application}

Three different examples are considered in this paper to test the proposed approach. As the first test case, a linear ordinary differential equation (ODE) problem, exponential growth, is presented. Then, the PIFI approach is applied to the 1D heat conduction equation, a nonlinear ODE problem. This test case is particularly important as it is the only one using a differential equation-based low-fidelity model. Lastly, the proposed approach is tested on a more complex problem, turbulent flow field modeling. This problem is governed by a 3D nonlinear partial differential equation (PDE), and hence, an important test case to evaluate the feasibility of the proposed approach in real engineering problems.

For every example, the field inversion stage is performed for three different scenarios. First, a Full Data Field Inversion (Full FI) is performed by using the observation for every grid point, which is in line with traditional field inversion applications. Then, sparse data is extracted within the full observation set and it is employed to run the only data-driven field inversion, denoted as Reduced Data FI (Reduced FI) in the rest of the paper. This scenario is run to illustrate the limitations of the conventional approach. Lastly, the same data loss used in the Reduced FI is employed by also incorporating the appropriate physical loss. This latest scenario is denoted as Physics-Informed Field Inversion (PIFI). The comparison of these three cases enables observing the advantage obtained from the physics-informed approach.

All inversion phases are performed by using the BFGS \cite{bfgs} implementation from the SciPy package \cite{scipy}, and sensitivities are calculated by using the automatic differentiation feature of JAX.

\subsection{Exponential Growth (Linear ODE)}
\label{subsec:ode}

The first and simplest example considered in this study is exponential growth, a well-known ordinary differential equation. The governing equation for this case is given in Equation \ref{eq:f_ode}.

\begin{equation}
    f(x, y) = \frac{dy}{dx} - ky = 0
    \label{eq:f_ode}
\end{equation}

In this equation, $y$ is the output state, $x$ is the input state which is typically time, and $k$ is the growth rate constant. In this example, $k$ is considered as a known and it is set to $5.0$. The 1D solution space is defined as $x \in [0.0, \: 1.0]$.

\subsubsection{Observations}
\label{subsubsec:ode_observations}

An analytical solution is available for the exponential growth equation, and it is given in Equation \ref{eq:M_ode}. This analytical solution can be employed as the observations of this example which are generated by using 20 evenly distributed points in the input space.

\begin{equation}
    \mathcal{M}(x) = e^{kx}
    \label{eq:M_ode}
\end{equation}

\subsubsection{Low-Fidelity Model}
\label{subsubsec:ode_lofi_model}

A dummy model is introduced to be employed as the low-fidelity model for this example. This simple model approximates the solution as a line with the slope of $m=100.0$, as defined in Equation \ref{eq:D_ode}.

\begin{equation}
    \mathcal{D}(x) = m x
    \label{eq:D_ode}
\end{equation}

To enhance the prediction capability of this dummy model, a correction term, $\beta$, is introduced as a function of $x$. This correlation will be applied as the multiplier to the slope term of the equation. The new low-fidelity model with the introduced correction term is given in Equation \ref{eq:H_ode}.

\begin{equation}
    \mathcal{H}_\beta(x) = \beta(x) m x
    \label{eq:H_ode}
\end{equation}

\subsubsection{Results}
\label{subsubsec:ode_results}

% Summary of the scenarios
Using the methodology explained in Section \ref{sec:methodology}, three distinct field inversion scenarios are performed to find their corresponding optimal correction distributions. The first case, Full FI, uses all 20 evenly distributed observations across the entire input space, while the reduced data cases, Reduced FI and PIFI, use only the data obtained from the first half of the observations, as summarized in Table \ref{tbl:ode_summary}.

\begin{table}[htbp]
\centering
\caption{Summary of Field Inversion Scenarios for the Linear ODE Example}
\resizebox{\textwidth}{!}{
\begin{tabular}{l|ccc}
    \toprule
    & \textbf{Data Loss} & \textbf{Reg. Loss} & \textbf{Phys. Loss} \\
    \midrule
    \textbf{Full FI} & All observations & Yes & No \\
    \textbf{Red. FI} & \makecell{Only first half of the observations} & Yes & No \\
    \textbf{PIFI} & \makecell{Only first half of the observations} & Yes &  Yes \\
    \bottomrule
\end{tabular}}
\label{tbl:ode_summary}
\end{table}

% Hyperparameter search study
Before comparing the results for these three different scenarios, a hyperparameter study is conducted to find the optimal loss weights for the PIFI solution. During this study, the data loss weight is fixed such that $\alpha_\text{reg}=1.0$, while other loss weights, $\alpha_\text{reg}$ and $\alpha_\text{phys}$, are under investigation. A grid search is performed to find the loss weights that yield the highest R$^2$ score. This search also provides valuable insight into the sensitivity of the proposed approach for the chosen loss weights.

The search results are presented in a heat map format in Table \ref{tbl:ode_parameter_search}. The highest fitting score is obtained by the combination of $\alpha_\text{reg}=1.0 \times 10^{-4}$ and  $\alpha_\text{phys}=1.0 \times 10^{-2}$, achieving an R$^2$ score of 0.9994, which is very close to a perfect recovery. It is observed that the accuracy sensitivity primarily depends on the physical loss weight, with a very limited effect from the prior distribution regularization weight. As the physical loss weight decreases, the PIFI results become closer to the Reduced FI result and exhibit lower recovery performance since the advantage of physics-informed inversion diminishes. On the other hand, very large $\alpha_\text{phys}$ values also lead to poor performance of the PIFI model as the impact of the data loss gets relatively smaller. Nonetheless, the search results show that the acceptable accuracy range for PIFI application is quite broad, varying from at least $1.0 \times 10^{-4}$ to $1.0$. This provides a good margin for selecting hyperparameters for the PIFI in this example.

\begin{table}[htbp]
\centering
\caption{R$^2$ Score Distribution w.r.t. Loss Weights for the Linear ODE Example ($\alpha_\text{data}=1.0$)}
\resizebox{\textwidth}{!}{
\begin{tabular}{|c|c|c|c|c|c|c|}
\hline
\rowcolor{label_color}
\backslashbox{$\mathbf{\alpha_{reg}}$}{$\mathbf{\alpha_{phys}}$} & \textbf{1.0e+02} & \textbf{1.0e+00} & \textbf{1.0e-02} & \textbf{1.0e-04} & \textbf{1.0e-06} & \textbf{1.0e-08} \\ \hline
\cellcolor{label_color}\textbf{1.0e+00} & \cellcolor[rgb]{0.8129585389996137, 0.9010309068633383, 0.8128886843791447}0.7430 & \cellcolor[rgb]{0.4700051728521834, 0.7579652275736392, 0.4699922497473967}0.9515 & \cellcolor[rgb]{0.39426087112637054, 0.7263678996323559, 0.3942605218532682}0.9981 & \cellcolor[rgb]{0.8318946144310668, 0.9089302388486591, 0.8318216163526769}0.7318 & \cellcolor[rgb]{0.9286789999696055, 0.9493046023291877, 0.9285899353285076}0.6731 & \cellcolor[rgb]{0.9286789999696055, 0.9493046023291877, 0.9285899353285076}0.6720 \\ \hline
\cellcolor{label_color}\textbf{1.0e-02} & \cellcolor[rgb]{0.8129585389996137, 0.9010309068633383, 0.8128886843791447}0.7433 & \cellcolor[rgb]{0.5162933572401802, 0.7772747057599789, 0.5162727501271418}0.9238 & \cellcolor[rgb]{0.39215686274509803, 0.7254901960784313, 0.39215686274509803}0.9991 & \cellcolor[rgb]{0.40057289627018833, 0.7290010102941296, 0.4005714991777789}0.9934 & \cellcolor[rgb]{0.8339986228123395, 0.9098079424025837, 0.8339252754608472}0.7304 & \cellcolor[rgb]{0.9286789999696055, 0.9493046023291877, 0.9285899353285076}0.6720 \\ \hline
\cellcolor{label_color}\textbf{1.0e-04} & \cellcolor[rgb]{0.8129585389996137, 0.9010309068633383, 0.8128886843791447}0.7433 & \cellcolor[rgb]{0.5120853404776349, 0.7755192986521298, 0.5120654319108013}0.9263 & \cellcolor[rgb]{0.39215686274509803, 0.7254901960784313, 0.39215686274509803}\textbf{0.9994} & \cellcolor[rgb]{0.40057289627018833, 0.7290010102941296, 0.4005714991777789}0.9931 & \cellcolor[rgb]{0.8276865976685217, 0.90717483174081, 0.8276142981363364}0.7343 & \cellcolor[rgb]{0.9286789999696055, 0.9493046023291877, 0.9285899353285076}0.6720 \\ \hline
\cellcolor{label_color}\textbf{1.0e-06} & \cellcolor[rgb]{0.8129585389996137, 0.9010309068633383, 0.8128886843791447}0.7433 & \cellcolor[rgb]{0.5036693069525446, 0.7720084844364317, 0.5036507954781204}0.9307 & \cellcolor[rgb]{0.39215686274509803, 0.7254901960784313, 0.39215686274509803}0.9989 & \cellcolor[rgb]{0.40057289627018833, 0.7290010102941296, 0.4005714991777789}0.9936 & \cellcolor[rgb]{0.8276865976685217, 0.90717483174081, 0.8276142981363364}0.7345 & \cellcolor[rgb]{0.9286789999696055, 0.9493046023291877, 0.9285899353285076}0.6720 \\ \hline
\cellcolor{label_color}\textbf{1.0e-08} & \cellcolor[rgb]{0.8129585389996137, 0.9010309068633383, 0.8128886843791447}0.7433 & \cellcolor[rgb]{0.5120853404776349, 0.7755192986521298, 0.5120654319108013}0.9253 & \cellcolor[rgb]{0.39215686274509803, 0.7254901960784313, 0.39215686274509803}0.9991 & \cellcolor[rgb]{0.40057289627018833, 0.7290010102941296, 0.4005714991777789}0.9931 & \cellcolor[rgb]{0.8276865976685217, 0.90717483174081, 0.8276142981363364}0.7346 & \cellcolor[rgb]{0.9286789999696055, 0.9493046023291877, 0.9285899353285076}0.6720 \\ \hline
\cellcolor{label_color}\textbf{1.0e-10} & \cellcolor[rgb]{0.8129585389996137, 0.9010309068633383, 0.8128886843791447}0.7433 & \cellcolor[rgb]{0.5183973656214527, 0.7781524093139034, 0.518376409235312}0.9224 & \cellcolor[rgb]{0.39215686274509803, 0.7254901960784313, 0.39215686274509803}0.9991 & \cellcolor[rgb]{0.40057289627018833, 0.7290010102941296, 0.4005714991777789}0.9933 & \cellcolor[rgb]{0.8276865976685217, 0.90717483174081, 0.8276142981363364}0.7346 & \cellcolor[rgb]{0.9286789999696055, 0.9493046023291877, 0.9285899353285076}0.6720 \\ \hline
\end{tabular}

}
\label{tbl:ode_parameter_search}
\end{table}

\begin{figure}[htbp]
    \begin{center}
        \includegraphics[width=0.6\textwidth, trim=5 10 5 5, clip]{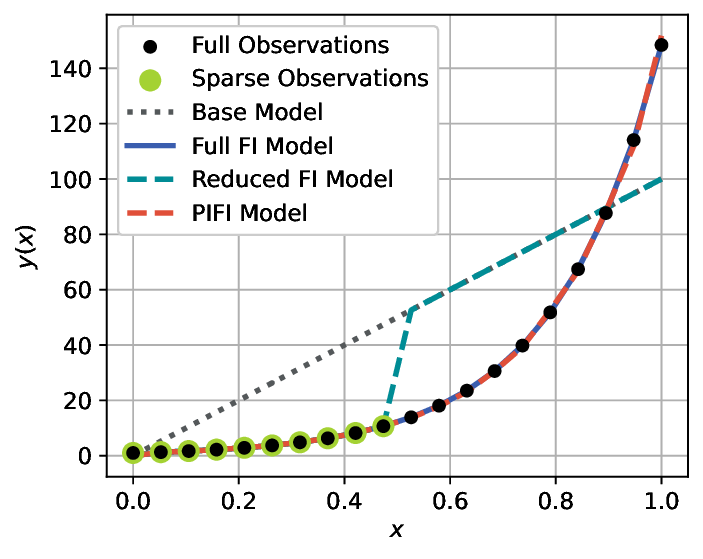}
    \caption{ Exponential Growth Case Results for k=5: Reduced FI using only first half of the observations follows the baseline results at the unseen region, whereas PIFI gives perfect agreement with the true results at that region as well.}
    \label{fig:ode_result}
    \end{center}
\end{figure}

% Results for the optimal case
PIFI results obtained using the optimal loss weights are plotted in Figure \ref{fig:ode_result}, along with the Base, Full FI, and Reduced FI model solutions. The same $\alpha_\text{reg}$ value used in the PIFI ($1.0 \times 10^{-4}$) is also applied to the loss functions for the Full and Reduced FI cases.

% Full FI
The Full FI model uses all of the available observations to obtain the optimal correction distribution. The results obtained from this case show that our low-fidelity model can accurately recover the observations if the appropriate correction set, $\beta(x)$, is used. This perfect matching represents the upper limit that can be achieved after a field inversion phase, and the goal of our proposed approach is to obtain a similar accuracy using a limited number of data points. 

\begin{figure}[htbp]
    \begin{center}
        \includegraphics[width=0.6\textwidth, trim=5 10 5 5, clip]{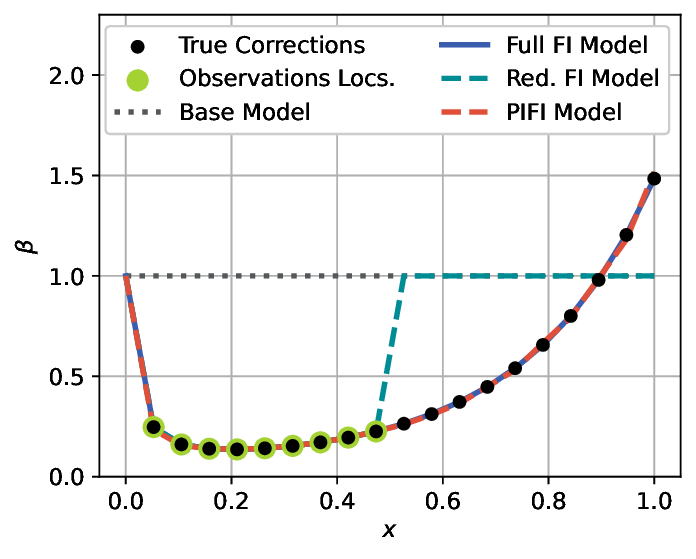}
    \caption{Distributions of Correction Parameters for the Exponential Growth Case: PIFI closely matches the distribution obtained from the Full FI, whereas the Reduced FI fails to correct the second half of the domain, where we assume no data is available.}
    \label{fig:ode_betas}
    \end{center}
\end{figure}

% Red FI
If we apply the same field inversion framework for the case where only the first half of the observations are known, we obtain the Reduced FI case. The model corrected after this inversion can accurately recover the first half of the observations, i.e., the seen points. However, it fails to correct the predictions for the second half of the solution domain and basically reverts to the baseline model due to the lack of gradient information at that region during the inverse optimization stage. This issue can be also seen from the correction distribution of the case presented in Figure \ref{fig:ode_betas} where the prior distribution is preserved in the second half of the corrections.

% PIFI
To address this issue, a physical loss term is also considered, and the PIFI results are obtained. This loss enables us to obtain gradient information not only for the known data points but also for the unseen ones. The effect of this gradient information is visible in Figure \ref{fig:ode_betas} where the optimal correction obtained from PIFI does not return the prior value for the second half of the solution space, unlike the Reduced FI case. In fact, it gives an almost identical correction parameter distribution to the one obtained from the Full FI model. This close alignment also results in a perfect recovery of the solution as shown in Figure \ref{fig:ode_result}. This success in the truncated data example is particularly important for evaluating the extrapolation capability of the PIFI approach.

\begin{figure}[htbp]
    \begin{center}
        \includegraphics[width=0.6\textwidth, trim=5 10 5 5, clip]{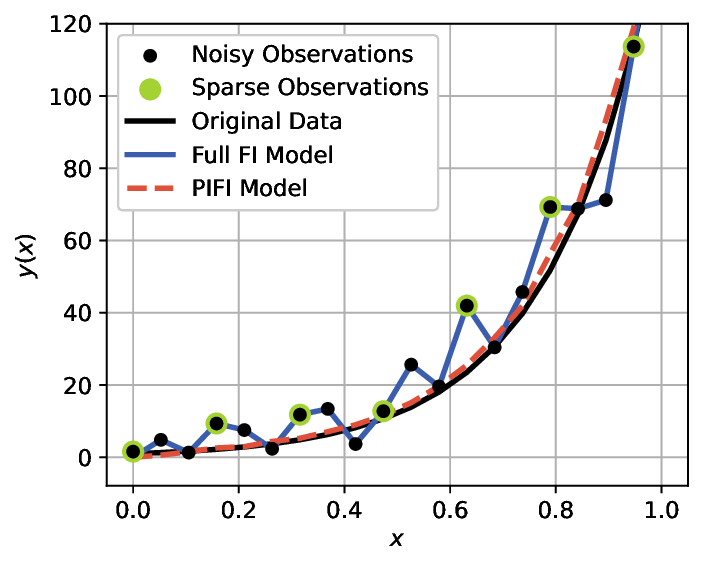}
    \caption{ Exponential Growth Case Results for k=5 with Noisy Observations: PIFI gives very close results to the original data by filtering the unphysical noise of the observations.}
    \label{fig:ode_with_noise_result}
    \end{center}
\end{figure}

% Noisy data
Although we have noted that increasing the weight of the physical loss beyond its optimal value decreases the accuracy of the PIFI solution, this threshold can be significantly higher for noisy observations. To test this claim, we slightly modified the ODE example. In this version, random noise is added to the original data for the observations. Additionally, evenly distributed sparse data is extracted from the observations to be used instead of the truncated data. The loss weights, $\alpha_\text{reg}$ is kept the same at $1.0 \times 10^{-4}$, but the $\alpha_\text{phys}$ value is increased to $1.0 \times 10^{1}$. The PIFI result obtained for this case is compared with the Full FI model in Figure \ref{fig:ode_with_noise_result}. In this example, the Full FI model is not the best choice as it inherently overfits the noisy data provided. When PIFI is applied with a greater physical loss weight, the framework avoids overfitting the provided data by acting as a form of adaptive regularization. It uses the information coming from the known data but prioritizes the physical loss, and results in filtered predictions that are very close to the original data without noise.

%%%%%%%%%%%%%%%%%%%%%%%%%%%%%%%%%%%%%%%%%%%%%%%%%%%%%%%%%%%%%%%%%%%%%%%%%%%%%%%%%%%%%%%%%%%%%%%%%%%%

\subsection{1D Heat Conduction (Nonlinear ODE)}
\label{subsec:nonlinear_ode}

The example in this section is the one-dimensional heat conduction equation with radiative and convective heat sources, which is one of the test cases of Parish and Duraisamy \cite{Parish_2016}. This example is also important in the content of this paper as it is the only example using a differential equation-based low-fidelity model, which transfers information between grid points, unlike the analytical models. The governing equation of the problem is defined in Equation \ref{eq:f_nonlinear_ode}.
\begin{equation}
    f(z, T) = \frac{d^2T}{dz^2} - \epsilon(T) (T^4-T_{\infty}^4) - h(T-T_\infty) = 0
    \label{eq:f_nonlinear_ode}
\end{equation}

In this example, the convection coefficient, $h$ is taken to be a constant, whereas the emissivity, $\epsilon$ is modeled as in Equation \ref{eq:emissivity}. The only difference in our implementation compared to the prior study is that the noise term in the emissivity equation is removed as we are working on a single realization.
\begin{equation}
    \epsilon(T)= \left(1 + 5 \sin \left( \frac{3 \pi}{200} T \right) + \exp \left(0.02T \right) \right) \times 10^{-4}
    \label{eq:emissivity}
\end{equation}

\subsubsection{Observations}
\label{subsubsec:nonlinear_ode_observations}

The observations are obtained by solving Equation \ref{eq:f_nonlinear_ode} using second-order central differentiation and a least-squares solution, as shown in Equation \ref{eq:M_nonlinear_ode}. The entire process is implemented using the JAX library to enable automatic differentiation. The solution space is formed as $z \in [0.0, \: 1.0]$, which is uniformly discretized by using 33 grid points. The convection coefficient is set to $h=0.5$, and the freestream temperature is taken as $T_\infty=50 \: K$. This value is also utilized as the initial guess for the least-squares framework. The temperatures at the boundaries are fixed to $T(z=0)=T(z=1)=0$ throughout the analysis.

\begin{equation}
    \mathcal{M}(z) = \arg \min_{T}\left\|f(z, T) \right\|
    \label{eq:M_nonlinear_ode}
\end{equation}

\subsubsection{Low-Fidelity Model}
\label{subsubsec:nonlinear_ode_lofi_model}

The low-fidelity model employed in this study is the same model considered in \cite{Parish_2016}, which is an imperfect modeling of the governing physics by neglecting the convective effect and utilizing a constant emissivity as $\epsilon_0=5.0 \times 10^{-4}$. The solution of this nonlinear ODE is again calculated by using the least squares method, as shown in Equation \ref{eq:D_nonlinear_ode}.

\begin{equation}
    \mathcal{D}(z) = \arg \min_{T}\left\| \frac{d^2T}{dz^2} - \epsilon_0 (T^4-T_{\infty}^4) \right\|
    \label{eq:D_nonlinear_ode}
\end{equation}

The correction coefficient is applied to the constant emissivity term as a function of $z$ as given in Equation \ref{eq:H_nonlinear_ode}.

\begin{equation}
    \mathcal{H}_\beta(z) = \arg \min_{T}\left\| \frac{d^2T}{dz^2} - \beta(z) \epsilon_0 (T^4-T_{\infty}^4) \right\|
    \label{eq:H_nonlinear_ode}
\end{equation}

\subsubsection{Results}
\label{subsubsec:nonlinear_ode_results}

% Inversion summaries
Field inversions are again performed using the same three scenarios: Full FI, Reduced FI, and PIFI. The Full FI uses all available 33 observations across the solution space to recover the temperature field, whereas the Reduced FI and PIFI only use observations at evenly distributed 7 data points, including the boundary conditions, as summarized in Table \ref{tbl:nonlinear_ode_summary}. In the PIFI approach, the physical loss is utilized as the residual of the true governing equation of the problem provided in Equation \ref{eq:f_nonlinear_ode}. The loss weights are kept the same as the previous study such that $\alpha_\text{data}=1.0$, $\alpha_\text{phys}=1.0 \times 10^{-2}$ and $\alpha_\text{reg}=1.0 \times 10^{-4}$, as it is seen that the insight obtained from the previous hyperparameter search study is also valid for this case.

\begin{table}[htbp]
\centering
\caption{Summary of Field Inversion Scenarios for the Nonlinear ODE Example}
\resizebox{\textwidth}{!}{
\begin{tabular}{l|ccc}
    \toprule
    & \textbf{Data Loss} & \textbf{Reg. Loss} & \textbf{Phys. Loss} \\
    \midrule
    \textbf{Full FI} & All observations & Yes & No \\
    \textbf{Red. FI} & \makecell{Evenly Distributed 7 Points} & Yes & No \\
    \textbf{PIFI} & \makecell{Evenly Distributed 7 Points} & Yes &  Yes \\
    \bottomrule
\end{tabular}}
\label{tbl:nonlinear_ode_summary}
\end{table}

\begin{figure}[htbp]
    \begin{center}
        \includegraphics[width=0.6\textwidth, trim=5 10 5 5, clip]{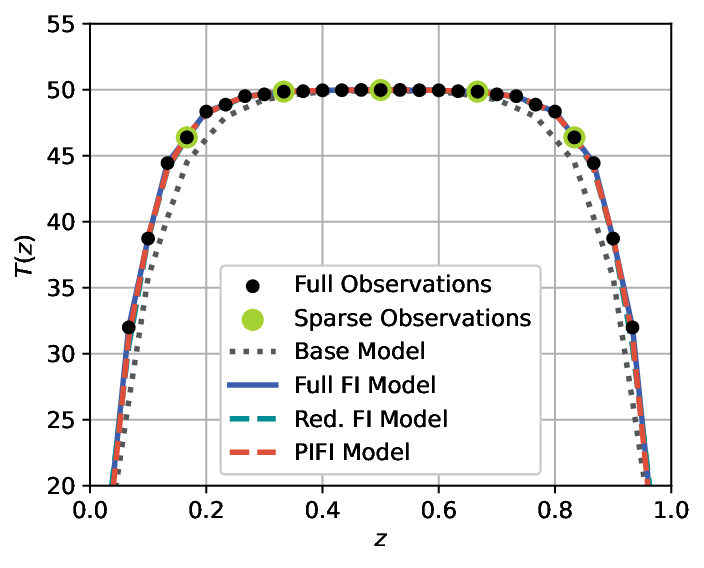}
    \caption{Recovery of Temperature Field for 1D Heat Conduction Case: All inversion scenarios recover the temperature field accurately.}
    \label{fig:nonlinear_ode_result}
    \end{center}
\end{figure}

% Recovery performance
For this example, the low-fidelity model already provides highly accurate results, given that the contribution from convection is limited compared to radiation. The recovery score of the baseline model is nearly 0.98, as given in Table \ref{tbl:r2_summary}, where the discrepancy is also visible in Figure \ref{fig:nonlinear_ode_result}. When the field inversion cases summarized in Table \ref{tbl:nonlinear_ode_summary} are performed for this case, almost perfect recoveries are obtained from every inversion case. This shows that the information-transferring nature of differential equation-based low-fidelity models allows for recovering observations, even at points where we do not have any observations. If the observations are dense enough, sparse data assimilation without physical loss may still be sufficient to perform a successful recovery, as observed in this case.

% Good recovery does not mean accurate posterior
However, accurate recovery performance does not always imply that the posterior correction parameter distribution is also accurate. This argument is the main discussion for this test case in \cite{Parish_2016} as well, where they showed that although recovery performances are highly accurate for every covariance matrix they have considered, the full covariance matrix calculation is still necessary to obtain accurate posterior statistics. To get this accurate covariance matrix, they utilized 100 realizations of this example case.

\begin{figure}[htbp]
    \begin{center}
        \includegraphics[width=0.6\textwidth, trim=5 10 5 5, clip]{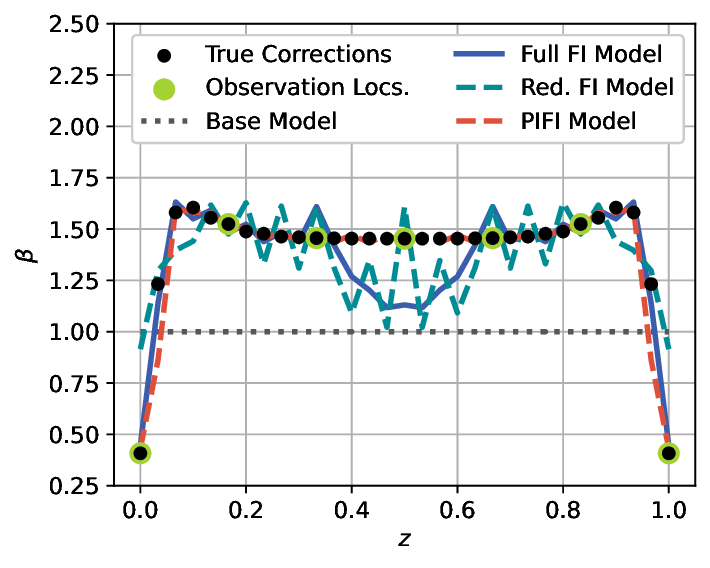}
    \caption{Correction Parameter Distributions for 1D Heat Conduction Case: Both Full FI and Reduced FI scenarios yield inaccurate posterior distributions due to the lack of prior knowledge, whereas PIFI matches the true distribution accurately.}
    \label{fig:nonlinear_ode_betas}
    \end{center}
\end{figure}

% Our posterior results - Full and Red FI
In this study, as we consider only one realization for each example during the field inversion stages, it is not possible to calculate such an accurate covariance matrix. As a result, even the Full FI case is not able to recover the analytical $\beta$ values, whose formula is derived in \cite{Parish_2016} and shown in Figure \ref{fig:nonlinear_ode_betas}. This inaccuracy also aligns with the findings of the reference study. When we consider the Reduced FI case with sparse observations, we see that even if the model can adjust its correction coefficients at points where observations are not available, these parameters oscillate more significantly compared to the Full FI case and do not match the analytical results again. This shows that although the sparse observations can be enough to recover the observations, the sparsity can still be problematic for obtaining accurate posterior correction coefficient distributions, which is also very important for the machine learning training stage.

% PIFI posterior
The main benefit of employing PIFI in this example appears at this point. The physical loss incorporation provides a posterior distribution that almost perfectly matches the true values, as shown in Figure \ref{fig:nonlinear_ode_betas}. This accurate result highlights the data efficiency of the proposed approach, where a single realization with limited observations is sufficient to provide an accurate posterior distribution. Therefore, for many applications, the PIFI approach can be more efficient than calculating a full covariance matrix by collecting a sufficient number of realizations.

%%%%%%%%%%%%%%%%%%%%%%%%%%%%%%%%%%%%%%%%%%%%%%%%%%%%%%%%%%%%%%%%%%%%%%%%%%%%%%%%%%%%%%%%%%%%

\subsection{Modeling Turbulent Flow Field (Nonlinear PDE)}
\label{subsec:complex_pde}

As the last example of this paper, a more complex case governed by a nonlinear PDE is considered. This example is particularly significant for evaluating the effectiveness of the approach in a more sophisticated problem that corresponds real-world engineering cases.

The case focuses on modeling the turbulence flow field for a jet flow scenario. Although FIML is predominantly used to correct the mean flow prediction of steady CFD analysis, such as correcting the turbulence closure models, by incorporating data obtained from higher-fidelity solvers, there are some cases where prediction of unsteady flow field statistics is also essential, such as in aeroacoustics applications. In a previous study \cite{Ugur_2024}, the authors employed FIML to enhance the prediction capabilities of a low-fidelity stochastic turbulent flow field generation model for both time-averaged and time-dependent statistics, which are essential inputs for an accurate farfield noise prediction framework. The example presented in this paper builds on that study, aiming for successful field inversion using sparse data. Although the framework is also capable of correcting time-dependent statistics, this example prioritizes the recovery of time-averaged turbulent velocity statistics to maintain conciseness, and only these results are presented accordingly.

This example also provides an opportunity to demonstrate another useful application of the proposed method such that the physical loss considered does not directly model the complete physics of the problem but rather focuses on a small portion of it. For instance, in this problem, the main governing equations are the nonlinear Navier-Stokes equations, which are computationally extremely expensive to solve. Even the high-fidelity data we have considered is an approximation to this governing physics. On the other hand, the low-fidelity model we are using is a pointwise model that is not even convection-aware. Billson et al. \cite{Billson_2003} demonstrated that applying some PDE-based filtering to make the generated turbulent flow field convection aware can be necessary to get accurate predictions from this low-fidelity model. However, this process increases the computation cost for each forward run and reduces the low cost advantage of the model. With this in mind, our objective in this example is to enhance the low-fidelity turbulence generation model by utilizing the 3D convection equation formulated in Equation \ref{eq:f_convection} to model the turbulent velocity component $\mathbf{u}'$. As we mentioned, this linear PDE loss does not capture the full physics of the problem but a small portion of it, which can be advantageous to decrease the cost of the inversion steps.

\begin{align}
    f(t, \bm x, \bm u') = \frac{\partial (\overline{\rho} \bm u')}{\partial t} + \frac{\partial (\overline{\rho \bm u} \cdot \bm u')}{\partial \bm x} = 0
    \label{eq:f_convection}
\end{align}

In Equation \ref{eq:f_convection}, the inputs, $t$ is time and $\mathbf{x}$ is the position vector, while $\overline{\rho}$ is mean density and $\overline{\rho \mathbf{u}}$ is mean momentum, obtained from steady flow solutions.

\subsubsection{Observations}
\label{subsubsec:sng_observations}

In this example, observations are obtained from a high-fidelity flow solver, specifically a Large Eddy Simulation (LES) solution of a round jet flow case. The flow under examination is a subsonic jet with a centerline velocity of $301.56 \: m/s$, its momentum thickness fraction is set as $0.108$ and the eddy viscosity is $0.001276$. For the analysis, a computational domain consisting of 1,521,230 elements is utilized. More detailed explanations about the jet geometry, computational grid, and analysis setup are available in \cite{Jivani_2021}.

To perform the LES analysis, an open-source CFD solver SU2 \cite{su2} is employed, and the simulations are conducted with a time step of $1.0 \times 10^{-5}s$. After completing the transient stage of the analysis, the statistics are averaged for 0.5 seconds.

\begin{equation}
    \mathcal{M}(\bm x, t) = \text{LES}(\bm x, t) 
\end{equation}

During the analysis, time-averaged fluctuating velocity components are recorded using moving averages for the whole domain. Additionally, to obtain time-dependent velocity statistics, 17 probes are positioned along the lipline of the nozzle. The locations of the probe points are presented in Figure \ref{fig:mean_flow_tke}.

\subsubsection{Low-Fidelity Model}
\label{subsubsec:sng_lofi_model}

As the lower-fidelity counterpart of this example, we employed a stochastic turbulent flow generation model called Stochastic Noise Generation (SNG) \cite{Bechara_1994, Bailly_1999} coupled with Reynolds Averaged Navier-Stokes (RANS). More specifically, RANS is employed to model the steady component of the flow field serving as an input for the SNG model. By using these pointwise inputs, SNG predicts time-dependent turbulent flow velocity components through an analytical approach. The main equation of SNG is based on a Fourier series approximation with a designated number of modes, $N$, as outlined in Equation \ref{eq:sng_main}. The theory of the model explained in this section is based on our previous study \cite{Ugur_2024}, which differs slightly from prior works due to certain modifications.

\begin{equation}
    \mathcal{D}(\bm x, t) = \bm u'(\bm x, t) = 2 \sum_{n=1}^N  \hat{u}_n \cos \left( \bm k_n \bm x + \psi_n + \omega_{0n} t \right) \bm \sigma_n
    \label{eq:sng_main}
\end{equation}

This equation takes the position vector $\bm x$, and time input to calculate the turbulent flow field vector $\bm u'$. The wave vector, $\bm k_n$, phase angle, $\psi_n$, and unit vector, $\bm \sigma_n$ are the random variables introducing the stochastic nature of the model, and their generation procedure is discussed in \cite{Bailly_1999}. The two deterministic parameters in this equation are the frequency term $\omega_n$ and the amplitude of the modes $\hat{u}_n$. 

In this study, the only term that is updated compared to the previous version is the calculation of $\omega_n$. This calculation is defined as a function of turbulent kinetic energy dissipation rate, $\epsilon$, obtained from RANS solution, as represented in Equation \ref{eq:sng_omega}. The reason for this choice is the better accuracy of the dissipation rate obtained from turbulence closure models compared with other candidate flow parameters.

\begin{equation}
    \omega_n = \sqrt[3]{\epsilon \| \bm k_n \|}
    \label{eq:sng_omega}
\end{equation}

The amplitude of the modes, $\hat{u}_n$ is defined by using energy integration, as shown in Equation \ref{eq:energy_integral} where the energy is term is modeled by the von Karman-Pao isotropic turbulence formula \cite{vonKarman1948progress} given in Equation \ref{eq:energy_model}.

\begin{equation}
    \hat{u}_n = \sqrt{E(k_n) \Delta k_n} \\
    \label{eq:energy_integral}
\end{equation}

\begin{equation}
    E(k) = \frac{2A}{3} \frac{K}{k_e} \left( \frac{k}{k_e} \right)^4 exp \left( -2 \left( \frac{k}{k_\eta}^2 \right) \right) \left( 1 + \left( \frac{k}{k_e} \right)^2 \right)^{(-17/6)} \\
    \label{eq:energy_model}
\end{equation}

The definitions of the terms standing in Equation \ref{eq:energy_model} are given in the following equations.

\begin{align}
    k_\eta &= \epsilon^{1/4} \nu^{-3/4} \\
    k_e &= 0.747/L_T \\
    L_T &= c_1 u'^3/\epsilon \label{eq:turb_length_scale}\\
    u' &= \sqrt{2K/3}
\end{align}

There are two parameters employed in these equations which are obtained from the RANS solution: turbulent kinetic energy, $K$ and energy dissipation rate, $\epsilon$. In addition, there are two free model parameters, $A$ and $c_1$, appearing in Equations \ref{eq:energy_model} and \ref{eq:turb_length_scale}. These free parameters are empirically calculated as 1.453 and 1.0, respectively, in previous studies \cite{Bailly_1999}. However, keeping them spatially constant does not necessarily gives the most accurate results, and tuning them can be necessary for different cases as discussed in \cite{Billson_2003} and \cite{Guglielmi_2025}. In this example, our aim is to correct these free parameters by applying spatially varying adjustments to their values used in the baseline model, as shown in Equation \ref{eq:beta_A_c1_corrections}.

\begin{align}
    A_{corr} = \beta_A(\bm x)  A  && {c_1}_{corr} = \beta_{c_1}(\bm x) c_1
    \label{eq:beta_A_c1_corrections}
\end{align}

In addition to these spatial corrections, applying a supplementary set of modal corrections is advantageous to correct the time-dependent behavior of the generated turbulent flow field \cite{Ugur_2024}. While the primary goal of this example is to correct the time-averaged turbulence statistics, such a modal correction is still desirable, as it provides the optimizer with sufficient degrees of freedom to enforce convection-aware results. Therefore, a set of modal corrections are applied to the $\hat{u}_n$ term through multiplication, as illustrated in Equation \ref{eq:beta_n_correction}.

\begin{align}
    {\hat{u}_{n,corr}} = \beta_{n}(n)  \hat{u}_n
    \label{eq:beta_n_correction}
\end{align}

By including the introduced corrections, the main equation of the SNG, given in Equation \ref{eq:sng_main}, can be rewritten as shown in Equation \ref{eq:corrected_sng}.

\begin{equation}
    \mathcal{H}_\beta(\bm x, t) = \bm u'(\bm x, t) = 2 \sum_{n=1}^N \beta_n(n) \hat{u}_n(\beta_A(\bm x), \beta_{c_1}(\bm x)) \cos \left( \bm k_n \bm x + \psi_n + \omega_{0n} t \right) \bm \sigma_n
    \label{eq:corrected_sng}
\end{equation}

As the first step to perform the low-fidelity model run, the steady flow solution is obtained using the RANS solver of SU2 with the SST $k-\omega$ turbulence model \cite{menter_SST_2003}. This closure model allows us to obtain the required inputs for the SNG: $K$ and $\epsilon$. The same grid used in the LES analysis is also employed to obtain this solution.

\begin{figure}[htbp]
    \begin{center}
        \includegraphics[width=0.9\textwidth, trim=5 5 5 5, clip]{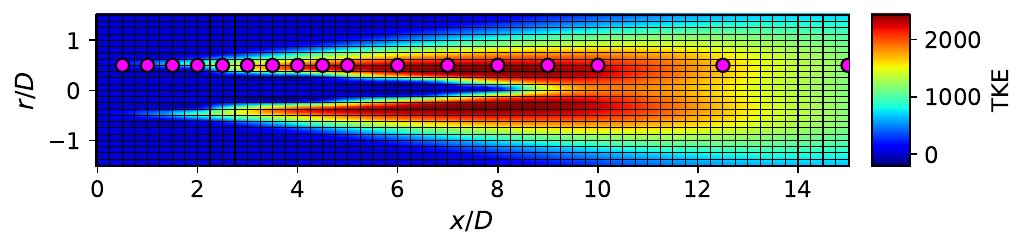}
        \caption{Illustration of the Computational Grid over the Turbulent Kinetic Energy of the Steady Flow Solution at the Meridian Plane: Magenta dots indicate the probe locations.}
        \label{fig:mean_flow_tke}
        \end{center}
\end{figure}

The SNG model is also implemented by using the JAX library to enable GPU usage and fast adjoint gradient calculations during the inversion stages. The inversion process is performed using a coarser, structured grid with dimensions of (61, 25, 25), located in the region of interest defined by ($x = [0, \: 15.0]$, $y = [-1.5, \: 1.5]$, $z = [-1.5, \: 1.5]$). The reason for the choice of this grid is that the computational grid used for CFD analyses is unnecessarily fine in some regions for the inversion process. Additionally, the structured grid enables faster and easier calculation of the gradients used in the physical loss calculation. The slice of the generated grid at the meridian plane is shown in Figure \ref{fig:mean_flow_tke}, presented with the turbulent kinetic energy solution obtained from RANS, which serves as the main input for the SNG model. Both LES and RANS solutions are interpolated onto this structured grid to be used in the SNG analysis. The duration of the analysis is set to 0.5 seconds to remain consistent with the observations. However, the time step is chosen to be greater with $2.5 \times 10^{-4} \: s$, as we also have the flexibility to use a more relaxed time discretization within this framework, as long as the desired frequency regime can still be analyzed. Finally, the wave number is set to vary linearly between 5 and 1000 in the SNG model, while 50 modes are employed for the Fourier series summation.

\subsubsection{Results}
\label{subsubsec:sng_results}

% rms loss
Let us first introduce some statistics and error measures used in this example. For the time-averaged statistics of the turbulent flow field, the root mean square (rms) values of the fluctuating velocity components are employed. To quantify the corresponding data loss, the MSE between the high-fidelity solution and low-fidelity SNG predictions is calculated, as shown in Equation \ref{eq:sng_rms_objective}, where $N_d$ denotes the number of data points where rms observations exist.

% PSD loss
Although the main aim of this example is to recover the spatial distribution of this time-averaged turbulent quantity using sparse data, some time-dependent statistics are also included in the data loss calculation, which can serve as a reference for the convection loss. Without providing any time-dependent data, it could be overly optimistic to expect effective incorporation of physical loss. This necessity also applies to time-dependent PIML studies. Therefore, another statistic associated with the time-dependent velocity measurements is employed as the power spectral density (PSD) of the fluctuating velocities. The PSD statistics are processed using Welch’s method \cite{Welch_1967} and Gaussian windowing \cite{Oppenheim_2014} with a standard deviation of 5 to obtain smoother results. The corresponding loss is again calculated using the MSE between LES and SNG. However, this time a logarithm is first applied to the PSD values, as shown in Equation \ref{eq:sng_psd_objective}. In this equation, $N_p$ denotes the number of probe locations used.

\begin{align}
    J_{\text{rms}} &= \frac{1}{3 N_d} {\sum_{k=1}^{N_d}} {\sum_{i=1}^3}  \left( {\text{rms}(u_i')}^k_{\text{SNG}} -  {\text{rms}(u_i')}^k_{\text{LES}}  \right)^2
    \label{eq:sng_rms_objective} \\
    J_{\text{PSD}} &= \frac{1}{3 N_p} {\sum_{k=1}^{N_p}} {\sum_{i=1}^3}  \left( \log(\text{PSD}(u_i'))^k_{\text{SNG}} -  \log(\text{PSD}(u_i'))^k_{\text{LES}} \right)^2
    \label{eq:sng_psd_objective}
\end{align}

% Combined data loss
Using two different sources for the data loss requires combining them with a weighted summation. Therefore, a weight of $0.05$ is selected and multiplied with the PSD loss, as shown in Equation \ref{eq:sng_data_loss}. This weight is chosen to prioritize rms recovery while still maintaining the PSD loss contribution at an effective level to guide the convection loss. To ensure a fair comparison between the three field inversion scenarios, the PSD data loss is applied in all cases with the same weight.

\begin{equation}
    J_{\text{data}} = J_{\text{rms}} + 0.05 \: J_{\text{PSD}} \label{eq:sng_data_loss}
\end{equation}

% FI case summaries
By using the data loss introduced, the same three field inversion scenarios are performed as summarized in Table \ref{tbl:complex_pde_summary}. The Full FI case uses the rms data from the whole flow field in addition to PSD values processed from the data collected from 17 probe locations located along the lipline of the nozzle. Reduced FI and PIFI cases, on the other hand, employ only the rms values and the PSD data at these specific probe locations without any information from other points by considering a similar setup to a typical experimental data collection procedure.

\begin{table}[htbp]
\centering
\caption{Summary of Field Inversion Scenarios for Turbulent Flow Modeling Case}
\resizebox{\textwidth}{!}{
\begin{tabular}{l|ccc}
    \toprule
    & \textbf{Data Loss} & \textbf{Reg. Loss} & \textbf{Phys. Loss} \\
    \midrule
    \textbf{Full FI} & \makecell{rms data for the whole flow field \& \\ PSD data at 17 probe locations} & Yes & No \\
    \textbf{Red. FI} & \makecell{rms and PSD data at 17 probe locations} & Yes & No \\
    \textbf{PIFI} & \makecell{rms and PSD data at 17 probe locations} & Yes &  Yes \\
    \bottomrule
\end{tabular}
}
\label{tbl:complex_pde_summary}
\end{table}

\begin{figure}[htbp]
    \begin{center}
        \includegraphics[width=0.6\textwidth, trim=5 5 5 5, clip]{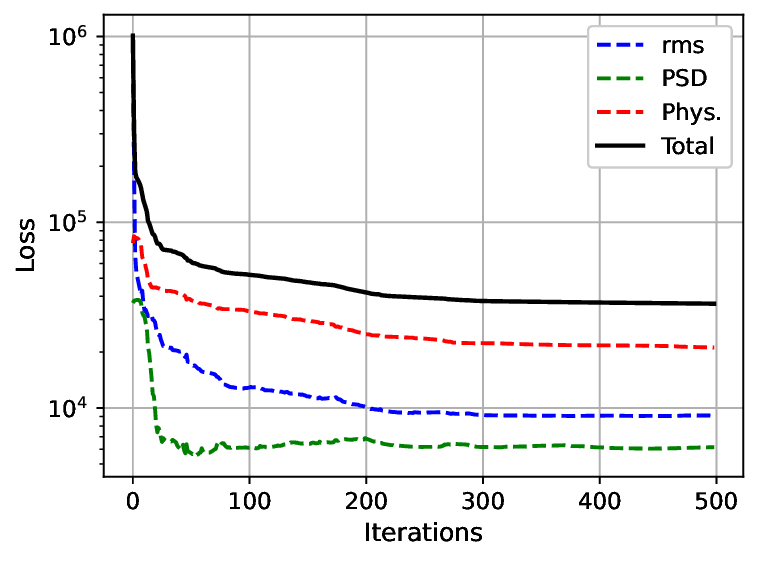}
        \caption{Inverse Optimization History for PIFI of the Jet Flow Case}
        \label{fig:pifi_convergence}
        \end{center}
\end{figure}

% Pysical loss and regulzarization weights and convergence tracking
In this example, the weight of the loss terms is not investigated through a hyperparameter search process due to the higher computational cost and the more complex inverse optimization stage of the problem. Instead, by using the conclusions drawn from previous cases as the starting point, a more expertise-involved hyperparameter tuning is performed. As a result, the weight of the physical loss term is set as $\alpha_\text{phys}=0.01$ while a smaller weight for the regularization term is chosen as $\alpha_\text{reg}=1.0 \times 10^{-5}$. In addition to the selected loss weights, a scaling factor of $1.0 \times 10^{4}$ is applied to all loss terms, since it provides a better optimum solution for this specific case. Furthermore, the convergence of the inversion processes is tracked in this example to stop the iterations after a sufficient number of iterations due to the higher cost of the optimization problem. The convergence history for the PIFI case is represented in Figure \ref{fig:pifi_convergence}. Regularization is not included in this graph, as its contribution is minimal compared to the other losses.

At the first iterations, it is seen that the most dominant term is the rms loss, which is the primary focus of the recovery process. Given the sparse data utilized in the PIFI process, the optimizer quickly adjusts the model parameters to fit the available rms data and substantially decreases the corresponding loss in the early iterations. This also results in a decrease in the PSD loss as it is not in a trade-off with the rms loss when they aim to recover the same data points. Similarly, the physical loss is not in a trade-off position with the other losses, as they contribute to generating a more physically turbulent flow field. After the first iterations, the greatest loss term becomes the physical loss and remains the most dominant contributor until the iterations are stopped after 500 iterations where the convergence is mostly satisfied. Throughout the optimization, the PSD loss was consistently the less effective contributor since it only serves as a guide for the physical loss, as mentioned earlier. For the studies aiming to achieve a higher accuracy in time-dependent statistics in the spectral domain, the weight of this loss term could be increased.

\begin{figure}[htbp]
    \centering\
    \captionsetup[subfigure]{oneside,margin={0.5cm,0cm}}
    \begin{subfigure}[b]{0.49\textwidth}
        \begin{center}
        \includegraphics[width=\textwidth, trim=5 5 5 5, clip]{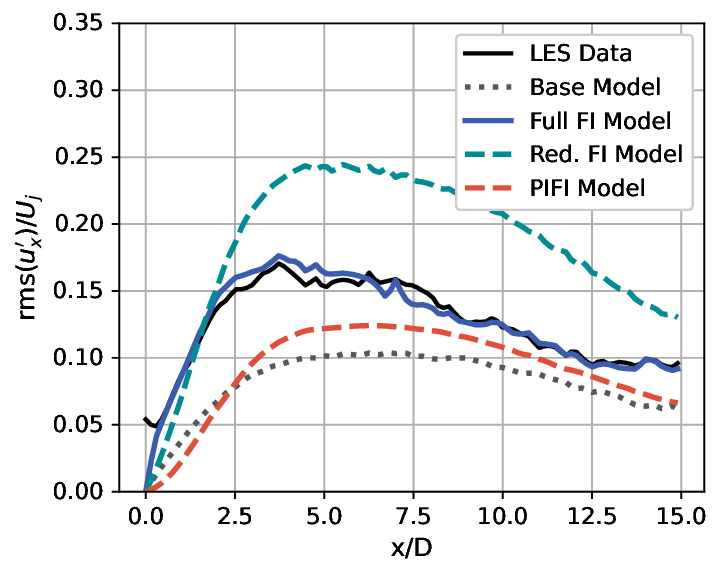}
        \caption{$r=-0.5D$}
        \label{fig:sng_rms_1}
        \end{center}
    \end{subfigure}
    \hfill
    \centering
    \begin{subfigure}[b]{0.49\textwidth}
        \begin{center}
        \includegraphics[width=\textwidth, trim=5 5 5 5, clip]{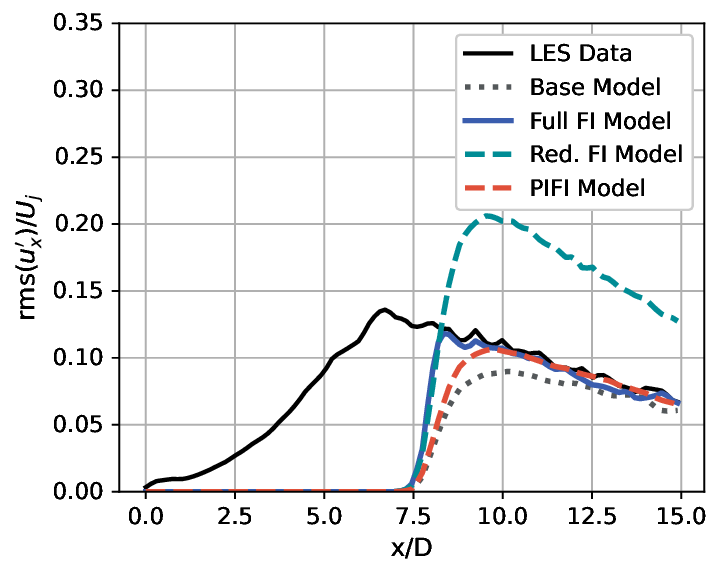}
        \caption{$r=0.0D$}
        \label{fig:sng_rms_2}
        \end{center}
    \end{subfigure}
    \begin{subfigure}[b]{0.49\textwidth}
        \begin{center}
        \includegraphics[width=\textwidth, trim=5 5 5 5, clip]{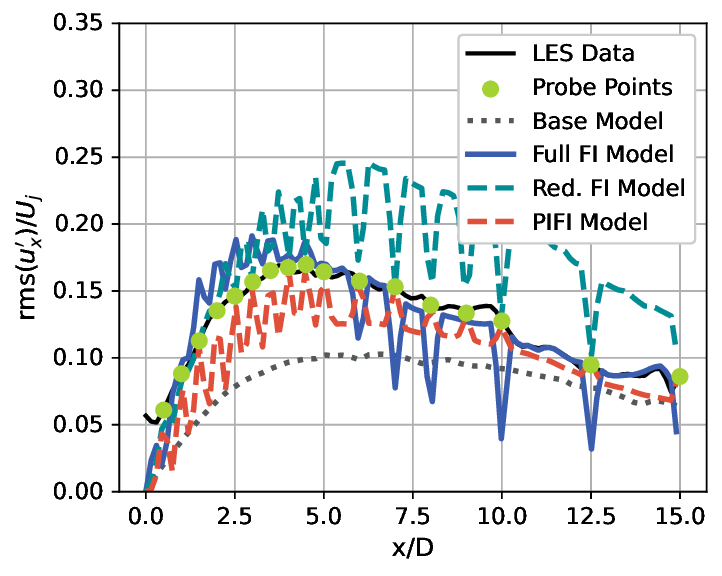}
        \caption{$r=0.5D$}
        \label{fig:sng_rms_3}
        \end{center}
    \end{subfigure}
    \hfill
    \begin{subfigure}[b]{0.49\textwidth}
        \begin{center}
        \includegraphics[width=\textwidth, trim=5 5 5 5, clip]{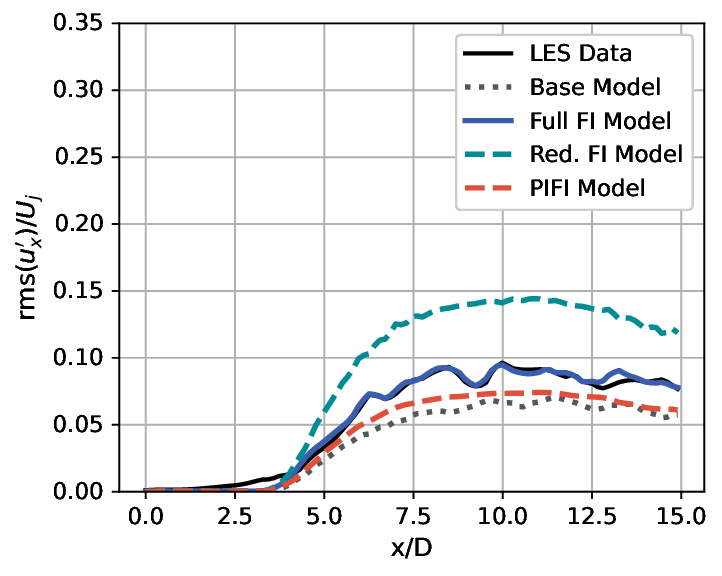}
        \caption{$r=1.0D$}
        \label{fig:sng_rms_4}
        \end{center}
    \end{subfigure}
    \caption{Distribution of rms Statistics of Streamwise Velocity Component Along the x-axis for Several Lines: PIFI accurately recovers the rms statistics at the probe locations and gives closer results to the LES data than Reduced FI and baseline models at other locations.}
    \label{fig:sng_rms_line}
\end{figure}

% Lower bound of the physical loss
It should be also noted that any physical loss in the PIFI approach potentially has a lower bound determined by the capabilities of the low-fidelity models, as they may not produce perfectly accurate physical solutions. For instance, in our jet case example, the results along the centerline of the nozzle, shown in Figure \ref{fig:sng_rms_line}, have a significant discrepancy in the nearfield of the nozzle, known as the potential core. This discrepancy is not essentially an error caused by the low-fidelity model but the input from the RANS solution, as the turbulent kinetic energy prediction of RANS is not accurate for jet flow cases, which is a known issue of RANS. Such restrictions can occur with any low-fidelity model, highlighting the importance of carefully choosing the physical loss weight, as very large weights can easily mislead the inversion process.

\begin{figure}[htbp]
    \centering
    \captionsetup[subfigure]{oneside,margin={-0.5cm,0cm}}
    \begin{subfigure}[h]{0.8\textwidth}
        \begin{center}
        \includegraphics[width=\textwidth, trim=5 5 5 5, clip]{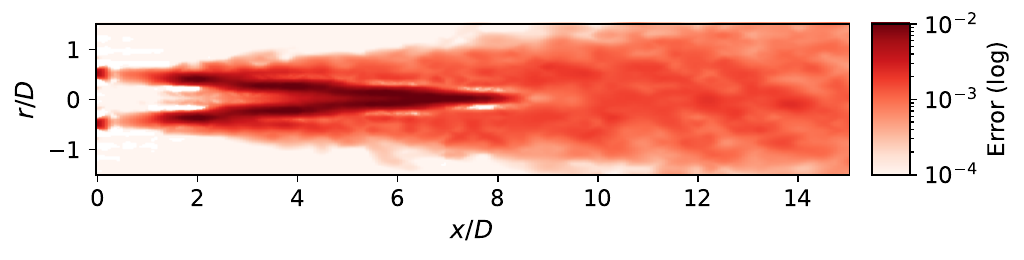}
        \caption{Base Model}
        \label{fig:base_rse}
        \end{center}
    \end{subfigure}
    \centering
    \begin{subfigure}[h]{0.8\textwidth}
        \begin{center}
        \includegraphics[width=\textwidth, trim=5 5 5 5, clip]{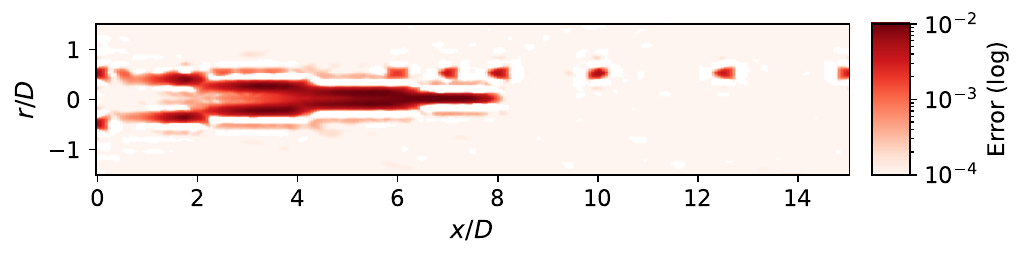}
        \caption{Full FI Model}
        \label{fig:full_rse}
        \end{center}
    \end{subfigure}
    \begin{subfigure}[h]{0.8\textwidth}
        \begin{center}
        \includegraphics[width=\textwidth, trim=5 5 5 5, clip]{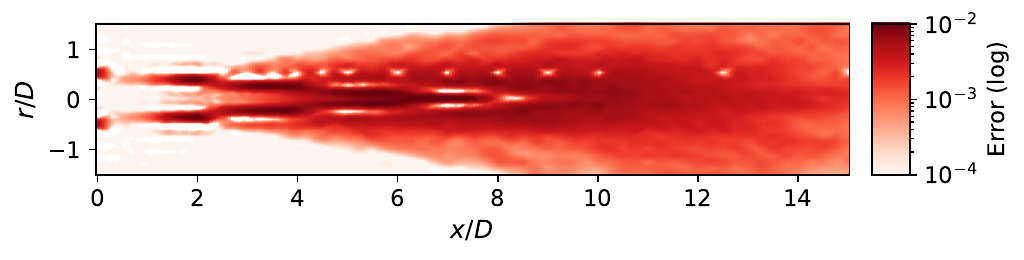}
        \caption{Reduced FI Model}
        \label{fig:red_rse}
        \end{center}
    \end{subfigure}
    \begin{subfigure}[h]{0.8\textwidth}
        \begin{center}
        \includegraphics[width=\textwidth, trim=5 5 5 5, clip]{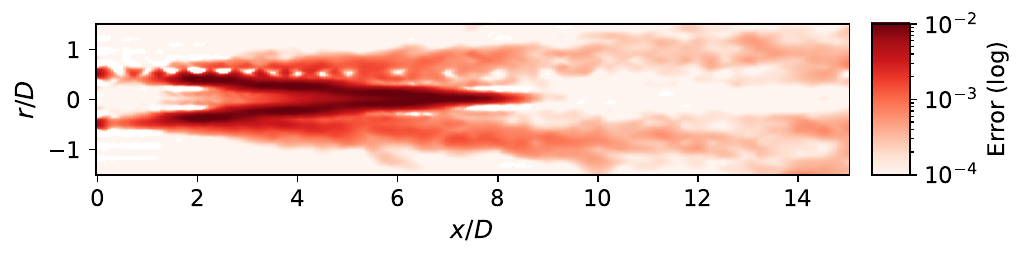}
        \caption{PIFI Model}
        \label{fig:pifi_rse}
        \end{center}
    \end{subfigure}
    \caption{Average of Root Squared Error of rms Components in the Meridian Plane for the Jet Flow Case: PIFI provides better recovery performance than Reduced FI and outperforms the baseline model.}
    \label{fig:rse_graph}
\end{figure}

% Results
Other field inversion scenarios, Full FI and Reduced FI are also performed with their corresponding data and loss weights by tracking convergence. Figure \ref{fig:rse_graph} represents the error in the meridian plane between the high-fidelity solution and the recovered turbulent flow field from the three field inversion cases, as well as the baseline model prediction. As the error measure, the root squared error of the rms values is calculated for each velocity component and the average of these three components is utilized. 

% Full FI
This comparison shows that the Full FI mostly recovers the LES data with high accuracy. This match is more clearly seen in Figure \ref{fig:sng_rms_line}, which shows the rms distribution of the streamwise velocity along several lines on the x-axis. This matching is almost perfect except for some regions, such as the potential core of the jet, which was previously stated as a known error due to the inaccurate RANS prediction at that region. Another obvious discrepancy between high-fidelity data and Full FI results is the jumps occurring at the probe locations. This behavior can seem strange at first since more information is provided to the optimizer for these points, and the recoveries of rms and PSD statistics are essentially correlated objectives. However, as the main objective of this scenario is matching rms values for the whole solution domain, it tries to adjust the modal corrections, having a global impact, to match both rms values of different points across the space and the PSD data provided for the probe locations. These two district objectives of the Full FI can cause a trade-off between PSD and rms recovery since some points on the solution domain can mislead the modal corrections for the sake of the rms recovery of these points, and result in some jumps at the probe locations. This highlights that even if the observations are available for the entire solution domain, a sparse optimization may still be beneficial for such a case to ensure focus on the most important regions of the flow field and prevent any misleading optimization process. 

\begin{figure}[htbp]
    \centering
    \captionsetup[subfigure]{oneside,margin={-0.5cm,0cm}}
    \begin{subfigure}[h]{0.8\textwidth}
        \begin{center}
        \includegraphics[width=\textwidth, trim=5 5 5 5, clip]{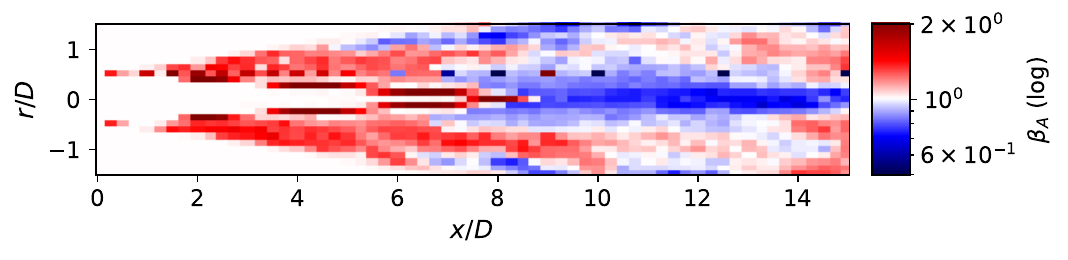}
        \caption{Full FI Model}
        \label{fig:full_beta_A}
        \end{center}
    \end{subfigure}
    \centering
    \begin{subfigure}[h]{0.8\textwidth}
        \begin{center}
        \includegraphics[width=\textwidth, trim=5 5 5 5, clip]{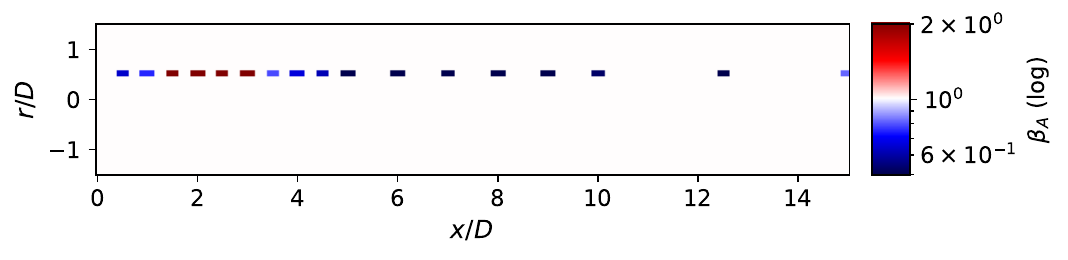}
        \caption{Reduced FI Model}
        \label{fig:red_beta_A}
        \end{center}
    \end{subfigure}
    \begin{subfigure}[h]{0.8\textwidth}
        \begin{center}
        \includegraphics[width=\textwidth, trim=5 5 5 5, clip]{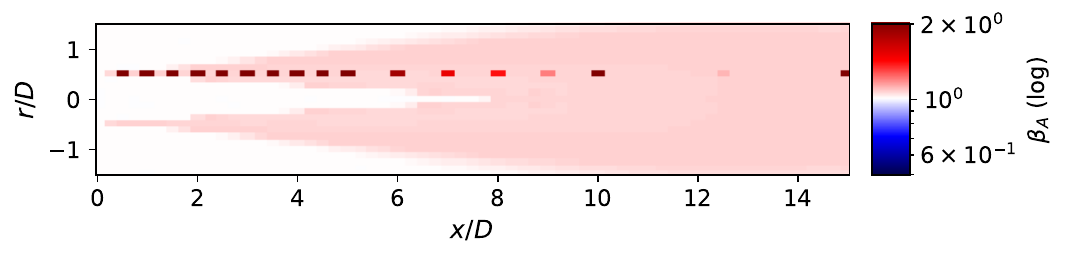}
        \caption{PIFI Model}
        \label{fig:pifi_beta_A}
        \end{center}
    \end{subfigure}
    \caption{Spatial Correction Distributions for the Turbulence Generation Model Parameter $A$: PIFI can adjust the correction parameters outside the probe points as well.}
    \label{fig:}
\end{figure}

\begin{figure}[htbp]
    \centering
    \captionsetup[subfigure]{oneside,margin={-0.5cm,0cm}}
    \begin{subfigure}[h]{0.8\textwidth}
        \begin{center}
        \includegraphics[width=\textwidth, trim=5 5 5 5, clip]{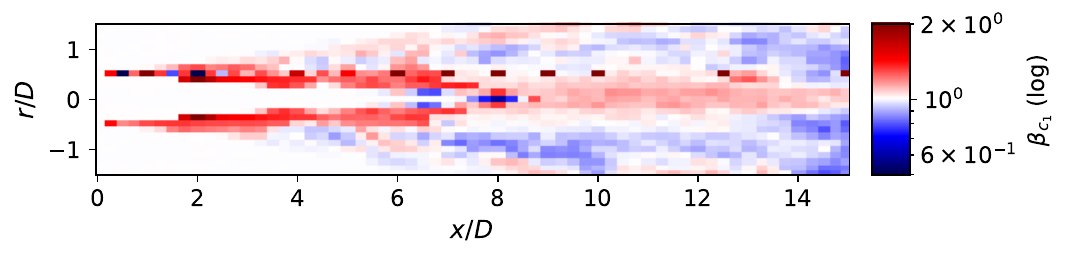}
        \caption{Full FI Model}
        \label{fig:full_beta_c1}
        \end{center}
    \end{subfigure}
    \centering
    \begin{subfigure}[h]{0.8\textwidth}
        \begin{center}
        \includegraphics[width=\textwidth, trim=5 5 5 5, clip]{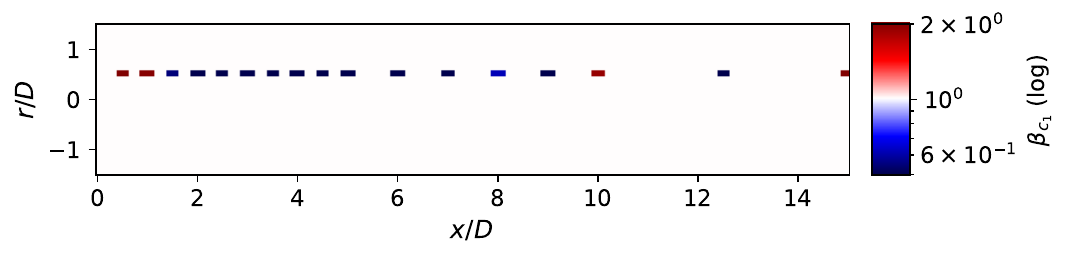}
        \caption{Reduced FI Model}
        \label{fig:red_beta_c1}
        \end{center}
    \end{subfigure}
    \begin{subfigure}[h]{0.8\textwidth}
        \begin{center}
        \includegraphics[width=\textwidth, trim=5 5 5 5, clip]{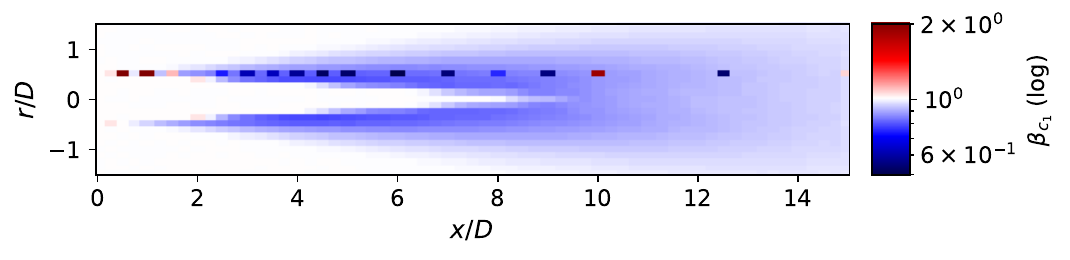}
        \caption{PIFI Model}
        \label{fig:pifi_beta_c1}
        \end{center}
    \end{subfigure}
    \caption{Spatial Correction Distributions for the Turbulence Generation Model Parameter $c_1$}
    \label{fig:}
\end{figure}

% Red FI
Although we have stated that a sparse field inversion can be desirable for this case, the Reduced FI results show that a direct sparse optimization may not be sufficient for such a purpose, as this approach tends to overfit the observed data points and ignores the other points along with their physical connections to the observations, given that the pointwise nature of the SNG model. The results in Figure \ref{fig:sng_rms_3} indicate that the Reduced FI can accurately recover the rms results at the probe locations. However, this causes an overfitting to the observed data and the model gives poor recovery performance for the rest of the flow field, which can be also seen in Figure \ref{fig:rse_graph}. This can be mainly attributed to the overfitting of modal corrections to the available data and not being able to adjust the spatial corrections outside of the probe locations, as shown in Figures \ref{fig:red_beta_A} and \ref{fig:red_beta_c1}.

% PIFI
PIFI can serve as a solution to this issue of the Reduced FI by incorporating the physical loss during the inversion stage. Figures \ref{fig:pifi_beta_A} and \ref{fig:pifi_beta_c1} show that PIFI is able to apply spatial corrections not only at the probe locations but the entire flow field. This is achieved thanks to the physical loss contribution which allows for calculating gradients for all points. It is important to note that the spatial correction distributions obtained from PIFI are not necessarily close to the Full FI corrections for this case. In fact, PIFI provides a more uniform correction distribution for the unseen points with a lower magnitude than the corrections at the observation locations. This smooth trend can even be desirable for noisy data cases where the physical loss can act as a form of regularization to avoid overfitting, similar to the noisy example we have considered in Section \ref{subsec:ode}. Also in this example, it is possible to see in Figure \ref{fig:sng_rms_line} that Full FI attempts to capture all details of the observed LES data whereas PIFI yields smoother results except for the probe locations. 

% PIFI benefit, computational time and R2
This dense gradient advantage of PIFI over the Reduced FI is also evident in the prediction results. Figures \ref{fig:rse_graph} and \ref{fig:sng_rms_line} highlight that PIFI not only recovers data provided at the probe locations but also improves the performance of the baseline model for the majority of the flow field. This improvement is especially more obvious at the downstream locations and the centerline of the nozzle. The superior recovery performance of PIFI over the Reduced FI and baseline models is also obvious from the R$^2$ scores as listed in Table \ref{tbl:r2_summary}, calculated by using rms values from all three velocity components. Although the recovery score of PIFI is lower than the 0.95 achieved by Full FI, it still has a promising score of 0.79 given the limited observations provided. As the final remark of this example, the convection calculation does not bring a significant time cost disadvantage thanks to the GPU-accelerated implementation. In fact, the inversion with the convection calculation increases the cost of adjoint gradient calculation cost by less than 3\% compared with the gradient calculation without this calculation, as shown in Table \ref{tbl:r2_summary}, making the use of this approach highly sensible for this case.

\section{Discussion \& Conclusion}
\label{sec:conclusion}

% Summary of the method
In this paper, a framework to correct the parameters of a low-fidelity model by incorporating both available data and the physical background of the problem is introduced. The physical knowledge is employed as a form of physical loss during the inverse optimization stage, in addition to the data loss, to determine the optimal low-fidelity model parameters. The physical loss is typically defined as the residual of the differential equation representing the physics of the problem. This addition is particularly useful for spatially correcting the parameters, as performed in the Field Inversion Machine Learning technique, for sparse observation cases such as experimentally obtained data. Incorporating a physical loss, similar to the Physics-Informed Machine Learning applications, can enable data assimilation with sparse observations by providing dense gradient information throughout the solution space.

% Examples and results
The three examples considered in this study showcase the capabilities of the proposed Physics-Informed Field Inversion framework. In the first test case, exponential growth linear ODE, we illustrated that PIFI can enhance the extrapolation capability of a low-fidelity model, and the physical loss can serve as an adaptive regularization to obtain more accurate recoveries from noisy observations. The 1D heat conduction case, governed by a nonlinear ODE is the second example covered, showcasing the impact of PIFI for the cases using differential equation-based low-fidelity models. Although the recovery performance of Reduced FI, which performs a solely data-driven inversion by using the sparse observations, is already accurate for this case, the posterior correction parameter distribution shows that PIFI provides a more accurate posterior distribution than the Full FI results using all observations of the realization. Finally, we presented the results for the turbulent velocity field generation case, governed by a complex 3D nonlinear PDE. This test case is important to assess the performance of PIFI in more sophisticated engineering cases. The results obtained from this test case show that PIFI can significantly improve the recovery performance compared with Reduced FI results for such a pointwise low-fidelity model. Although analytical models can be considered weak baselines for recovery with sparse observation purposes, they can be highly desirable for many applications, as the aim of developing low-fidelity models is essentially decreasing the cost as much as possible. Table \ref{tbl:r2_summary} summarizes the recovery capability of PIFI, comparing it with two field inversion scenarios.

\begin{table}[htbp]
\centering
\caption{Summary of R$^2$ Scores and the Time Cost Increase due to the Physical Loss Calculation for the Example Cases}
% \resizebox{\textwidth}{!}{
\begin{tabular}{l|ccc}
    \toprule
     & \textbf{\makecell{Exp. Growth \\ (Linear ODE)}} & \textbf{\makecell{Heat Conduction \\ (Nonlinear ODE)}}  & \textbf{\makecell{Turbulent Flow \\ (Nonlinear PDE)}}\\
    \midrule
    \textbf{Baseline} & 0.5148 & 0.9566 & 0.6645 \\
    \textbf{Full FI}  & 1.0000 & 1.0000 & 0.9519 \\
    \textbf{Red. FI}  & 0.6720 & 0.9990 & 0.3513 \\
    \textbf{PIFI}     & 0.9994 & 0.9996 & 0.7902 \\
    \midrule
    \textbf{Cost Inc.} & 55.82\% & 6.47\% & 2.93\% \\
    \bottomrule
\end{tabular}
% }
\label{tbl:r2_summary}
\end{table}

% PIFI vs PIML
An important remark about the proposed approach is its position within the scientific machine learning techniques. It cannot be considered a direct alternative to PIML approaches since the output states are not predicted by a universal approximator but rather by a low-fidelity model. It can be more accurate to classify this approach as an extension to the techniques performing model-level corrections. Such approaches require more expertise in the application domain to develop or employ a suitable low-fidelity model. This disables one of the key advantages of PIML applications, which is their flexibility to be applied to any case with minimal effort. On the other hand, the low-fidelity model-based approaches also offer a serious advantage over the ones using universal approximators, as they ensure the production of physically consistent solutions to some extent and generally maintain a strong baseline accuracy. These aspects are important for the reliability of the approach in both seen and unseen cases. Another difference between the proposed method and the PIML approaches is that the PIFI does not directly enable a mesh-free framework. Depending on the nature and implementation of the low-fidelity model employed, a computational mesh can be required for some cases to calculate the derivatives during the inversion stage, even if the low-fidelity model itself is mesh-free. This also eliminates an advantage of PIML which uses automatic differentiation features of machine learning models to calculate analytical derivatives.

% Computational cost discusiion
The main advantage of embedding the physical loss calculation into the inverse optimization process instead of employing a traditional differential equation solver is that the residual calculation of the differential equation is required only during the inversion stage in this approach. Once the model is corrected by using the extracted knowledge, the forward run cost of the enhanced low-fidelity model remains the same as the original low-fidelity model. Yet, a questionable aspect of the proposed approach can be its cost-benefit trade-off during the field inversion stage due to the physical loss calculation process. To address such a concern, we measured the time cost increase in the example cases caused by the addition of the physical loss calculation step, whose results are presented in Table \ref{tbl:r2_summary}. This increase is calculated as the time increment to perform one gradient calculation step, which is the most dominant cost in the optimization cycle. For cases using simpler low-fidelity models, such as exponential growth example, where the forward run time is relatively short, the impact of physical loss calculation becomes more evident. Our tests indicate that it can account for up to 56\% of the gradient calculation cost when the physical loss is not considered. However, when we consider the more complex turbulent flow generation case, or the heat conduction case employing a differential equation-based low-fidelity model, it is observed that the main cost is dominated by the forward run of the low-fidelity model, and the physical loss calculation only increases the cost by less than 3\% for the turbulent flow example. Additionally, for some cases of this example, it is seen that the physical loss addition can accelerate the convergence or reduce the number of iterations required in the line searches, resulting in shorter inversion process times overall.

% Data-efficiency for porterior statistics
One could also argue that a traditional solver could be used to feed the solely data-driven framework with comparable cost and accuracy instead of using this approach. However, Parish and Duraisamy \cite{Parish_2016} discussed the need for a full covariance matrix calculation to get accurate posterior statistics, which is as important as recovering the solution field for the FIML framework, as the posterior correction distribution serves as the feature set during the machine learning model training. This full covariance matrix calculation may require a high volume of samples for the problem considered. On the other hand, the PIFI approach can provide accurate posterior correction parameter distribution by using only one realization, as discussed in Section \ref{subsec:nonlinear_ode}. This can mitigate a drawback of solely data-driven field inversion by enabling data-efficient and accurate posterior statistics.

% Relax schemes
Furthermore, since the residual is calculated only as a loss but not to obtain the solution itself, PIFI provides us important flexibility in terms of the implementation of this residual. For example, in cases where the residual calculations are challenging and expensive, we can employ cheaper equations for the loss calculation, such as the turbulent velocity generation example presented in Section \ref{subsec:complex_pde}, where we employed the linear convection PDE as the physical loss, which is significantly cheaper and easier to implement than the nonlinear full Navier-Stokes equations. This flexibility also extends to using a coarser mesh and larger time steps, which can substantially reduce the computational cost. Additionally, using lower-order spatial and temporal discretizations is likely to harm the solution quality much less than in a traditional solver since there is no risk of cumulative error given that the solution is generated by the low-fidelity model. In fact, second-order spatial discretizations and first-order temporal discretizations, where applicable, are used in all examples of this paper, which were relaxed and easy schemes to implement. While implementing higher-order schemes may be necessary for some specific problems, we claim that the sensitivity of solution quality and accuracy to the PDE implementation in this approach is lower than that of traditional solvers.

% Conclusion and future work
The test cases we have considered in this paper demonstrate the benefits of the PIFI approach in different aspects. The cost incurred to achieve these enhancements is reasonable and becomes less consequential as the cost of the forward model increases. However, to more precisely discuss the advantages of PIFI, especially in terms of posterior statistics accuracy, the method should be tested further on a broader range of engineering cases combined with the machine learning stage of FIML. Additionally, the main drawback of the proposed approach can be considered as the necessity to know the governing physical equations, at least to some extent. While the general form of the equations is often known, some parameters may remain unknown in some cases. Such an issue can be addressed by combining the proposed approach with model discovery techniques, which is left as a follow-up study idea of this study.

\section*{Declaration of generative AI and AI-assisted technologies in the writing process}

During the preparation of this work the authors used AI tools (Grammarly AI \& ChatGPT) for grammatical corrections, spelling checks, and improving readability only. After using these tools, the authors reviewed and edited the content as needed and take full responsibility for the content of the published article.

\bibliographystyle{elsarticle-num} 
\bibliography{references}

\end{document}